\documentclass[preprint,prl,aps,amsmath,amssymb,superscriptaddress,floatfix,nobibnotes]{revtex4}

\pdfoutput=1 

\usepackage{graphicx}

\usepackage{dcolumn}

\usepackage{bm}

\usepackage{gensymb}                                                                                                                                                                                         

\usepackage{subfigure}

\usepackage{epsfig}

\usepackage{float}

\usepackage{enumerate}

\usepackage{tabularx} 

\usepackage[version=3]{mhchem}

\usepackage{color}

\usepackage{txfonts}

\usepackage{morefloats}

\mathchardef\mhyphen="2D

\graphicspath{{Plot/}{Bary/}{Sky/}}

\begin{document}


\rightline{DAMTP-2015-63}
\rightline{YITP-15-94}


\vskip 20pt

\title{\Large{Electromagnetic Transition Strengths for Light 

Nuclei in the Skyrme model}}

\vskip 20pt

\author{\bf {M. Haberichter}}%

 \email{m.haberichter@kent.ac.uk}

\affiliation{%
School of Mathematics, Statistics and Actuarial Science,\\ University of Kent, Canterbury CT2 7NF, U.K.\\
}%

\author{\bf{P. H. C. Lau}}%


 \email{pakhang.lau@yukawa.kyoto-u.ac.jp}

\affiliation{%
Yukawa Institute for Theoretical Physics, \\
 Kyoto University, Kitashirakawa Oiwakecho, Sakyo-ku, Kyoto City, Kyoto 606-8502, Japan\\
}%

\affiliation{%
Department of Applied Mathematics and Theoretical Physics, \\
 University of Cambridge, Wilberforce Road, Cambridge CB3 0WA, U.K.\\
}%

\author{\bf{N. S. Manton}}%

\email{N.S.Manton@damtp.cam.ac.uk}

\affiliation{%
Department of Applied Mathematics and Theoretical Physics, \\
University of Cambridge, Wilberforce Road, Cambridge CB3 0WA, U.K.\\
}%

\date{\today}


\vskip 20pt

\begin{abstract}

We calculate reduced $B(E2)$ electromagnetic transition strengths for light nuclei of mass numbers $B=8,12,16,20,24$ and $32$ within the Skyrme model. We find that the predicted transition strengths are of the correct order of magnitude and the computed intrinsic quadrupole moments match the experimentally observed effective nuclear shapes. For the Hoyle state we predict a large $B(E2)\!\uparrow$ value of  $0.0521\, \text{e}^2\text{b}^2$. For Oxygen-16, we can obtain a quantitative understanding of the ground state rotational band and the rotational excitations of the second spin-0 state, $0_2^+$.

\end{abstract}





\maketitle

\section{Introduction}

Radiative electromagnetic transitions between nuclear states are an excellent way to probe nuclear structure and to test nuclear structure models \cite{bohrMot1,bohr1998nuclear2,2014LNP...875...25J}. In even-even nuclei, the reduced transition probability $B(E2: 0_1^+\rightarrow 2_1 ^+)$  from the $0_1^+$ ground state to the first excited $2_1^+$ state is particularly important \cite{PhysRevC.46.164,Raman:1201zz}. $B(E2)$ transitions play a crucial role \cite{Raman:1201zz,Pritychenko:2013taa} in determining mean lifetimes of nuclear states, the nuclear potential deformation parameter $\beta$, the magnitude of intrinsic electric quadrupole moments, and the energy of low-lying levels of nuclei. Large quadrupole moments and transition strengths indicate collective effects in which many nucleons participate. 

The Skyrme model \cite{Skyrme:1961vq,Skyrme:1962vh} and its
topological soliton solutions (known as Skyrmions) have been found to
capture important features of light nuclei of even baryon number. As
in the $\alpha$-particle model of nuclei
\cite{blattWeiss,Brink1970143}, Skyrmions with topological charge $B$
a multiple of four are composed of charge four sub-units
\cite{Battye:2006na}. Here the role of the $\alpha$-particle is taken
by the cubic $B=4$ Skyrmion. The arrangements of $B=4$ cubes often
resemble \cite{Battye:2006na,Feist:2012ps} those discussed in the
$\alpha$-particle model. In addition, the allowed quantum states for
each Skyrmion of topological charge $B$ often match
\cite{Krusch:2002by,Krusch:2005iq,Manko:2007pr,Battye:2009ad,Lau:2014sva}
the ground and excited states of nuclei with mass number $B$ a
multiple of four. Among other successes of the Skyrme model is the
prediction of the excitation energy of states in the rotational bands
of Carbon-12, including the excitations of the Hoyle state
\cite{Lau:2014baa}. However, isoscalar quadrupole $E2$
transitions within the Skyrme model, which can provide
us with valuable information about the internal structure of nuclei,
and a non-trivial test of the model, have not yet been studied in
detail. Note that  isovector magnetic dipole $M1$
transitions have been discussed for Skyrmion states in the
literature. The $M1$ transition from a delta to a nucleon has been
considered \cite{Adkins:1983ya}, and also the transition from a
deuteron to its isovector state \cite{Braaten:1988cc}. In
appendix B of Ref.~\cite{Kopeliovich:2001yg} a connection between the
isovector magnetic moment operator and the Skyrmion's mixed inertia
tensor was established for arbitrary SU(2) Skyrmions. 

In the following, we briefly review the Skyrme model and its soliton solutions. For further details, we refer the interested reader to the literature \cite{Manton:2004tk,Brown:2009eh,Gisiger:1998tv}. The Skyrme model is a modified nonlinear sigma model, in which the sigma field and isotriplet of pion fields $\boldsymbol{\pi}$ are combined into an $S\!U(2)$-valued scalar field 
\begin{align}
U(\boldsymbol{x},t)=\sigma(\boldsymbol{x},t)1_2+i\boldsymbol{\pi}(\boldsymbol{x},t)\boldsymbol{\cdot\tau}\,,
\end{align}
where $\boldsymbol{\tau}$ denotes the triplet of Pauli matrices and the normalization constraint $\sigma^2+\boldsymbol{\pi\cdot\pi}=1$ is imposed. 

For a static Skyrme field $U(\boldsymbol{x})$, the energy in Skyrme units is
\begin{align}
E&=\int\left\{-\frac{1}{2}\text{Tr}\left(R_iR_i\right)-\frac{1}{16}\text{Tr}\left([R_i,R_j][R_i,R_j]\right)+m^2\text{Tr}\left(1_2- U\right)\right\}\text{d}^3x\,.
\label{Skyenergy_SU2}
\end{align}
Here, $R_i$ are the spatial components of the $S\!U(2)$-valued current $R_\mu=\left(\partial_\mu U\right)U^\dagger$, and  $m$ is a dimensionless pion mass parameter. Skyrme units are converted to physical energies and lengths (in MeV and fm) by the factors $F_\pi/4e_{\text{Sky}}$ and $2/e_{\text{Sky}}F_\pi$, respectively. $e_{\text{Sky}}$ is a dimensionless constant and $F_\pi$ can be interpreted as the pion decay constant. $m$ is related to the pion tree level mass $m_\pi$ via $m=2m_\pi/e_{\text{Sky}}F_\pi$. The energy and length conversion factors are fixed by comparison with experimental nuclear physics data. 

Skyrmions are critical points of the potential energy (\ref{Skyenergy_SU2}) and are characterized by a conserved, integer-valued topological charge
\begin{align}
B=-\frac{1}{24\pi^2}\int\,\epsilon_{ijk}\text{Tr}\left(R_iR_jR_k\right)\,\text{d}^3x\,.
\label{Sky_bary}
\end{align}
$B$ is the topological degree of the map $U:\mathbb{R}^3\rightarrow S\!U(2)$ at any given time, which is well-defined for fields satisfying the boundary conditions $\sigma\rightarrow 1$ and $\boldsymbol{\pi}\rightarrow \boldsymbol{0}$ as $|\boldsymbol{x}|\rightarrow\infty$. Physically, when semiclassically quantized \cite{Adkins:1983ya,Adkins:1983hy}, a Skyrmion of charge $B$ is interpreted as a nucleus of mass number (or baryon number) $B$. In nuclear physics, the notation for mass number is $A$ but we will keep our notation $B$ in this paper.

Skyrmion solutions with rescaled pion mass $m=1$ and with baryon number $B$ a multiple of four have been previously found \cite{Battye:2006na,Lau:2014baa,Lau:2014sva,Feist:2012ps}. For baryon numbers $B=8,\,12,\,16,\,20,\,24,\,32$, we recalculate the classical Skyrmion solutions using two different numerical relaxation techniques: nonlinear conjugate gradient \cite{NCG,Feist:2011aa,Feist:2012ps} and damped full field evolution \cite{Battye:2001qn}. Skyrmions are the solutions of minimal energy, or sometimes local minima or saddle points with energies close to minimal. Our calculations have led us to two new solutions with $B=24$. These are obtained by gluing together two copies of $B=12$ solutions. 

\begin{figure}[!htb]
\subfigure[]{\includegraphics[totalheight=3.cm]{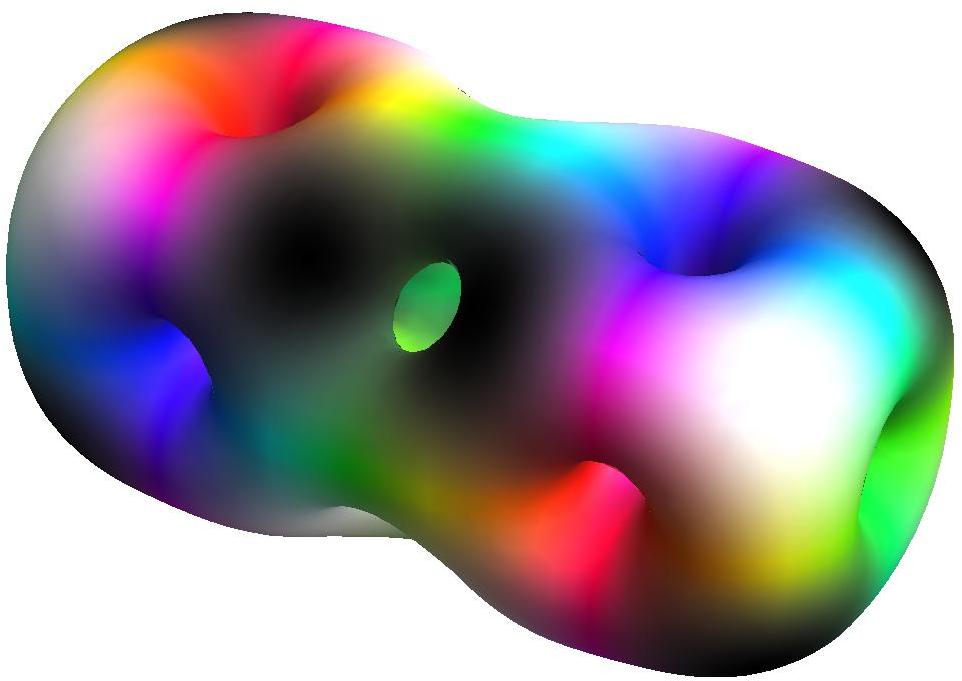}}\label{B8twist}
\subfigure[]{\includegraphics[totalheight=3.cm]{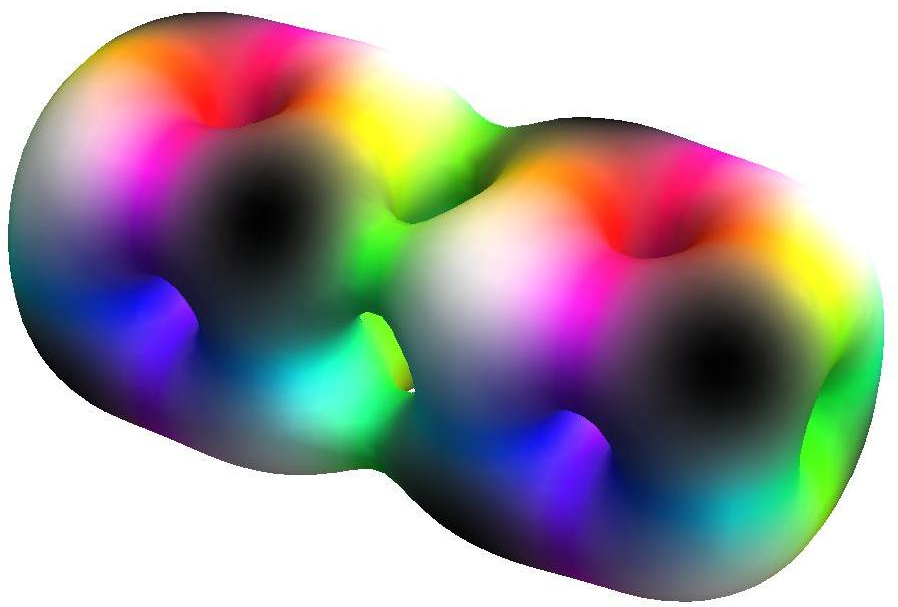}}\label{B8notwist}
\subfigure[]{\includegraphics[totalheight=3.cm]{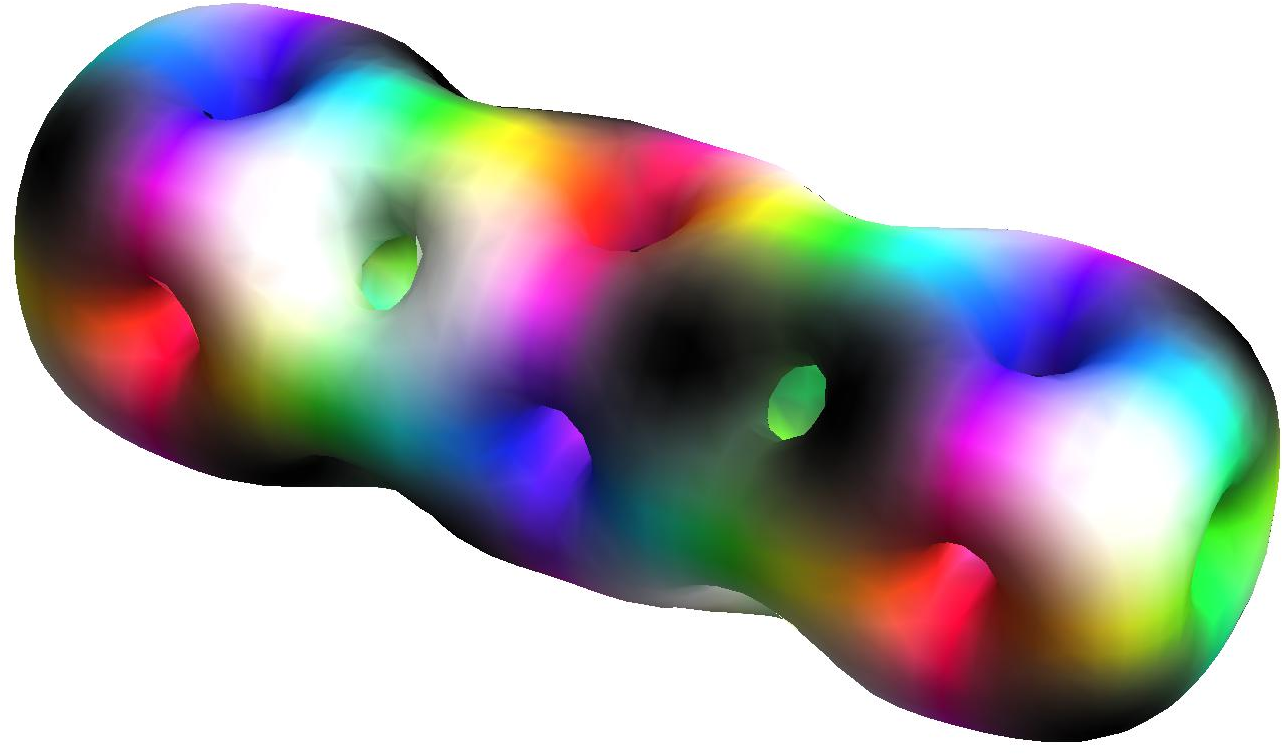}}\\
\subfigure[]{\includegraphics[totalheight=4.cm]{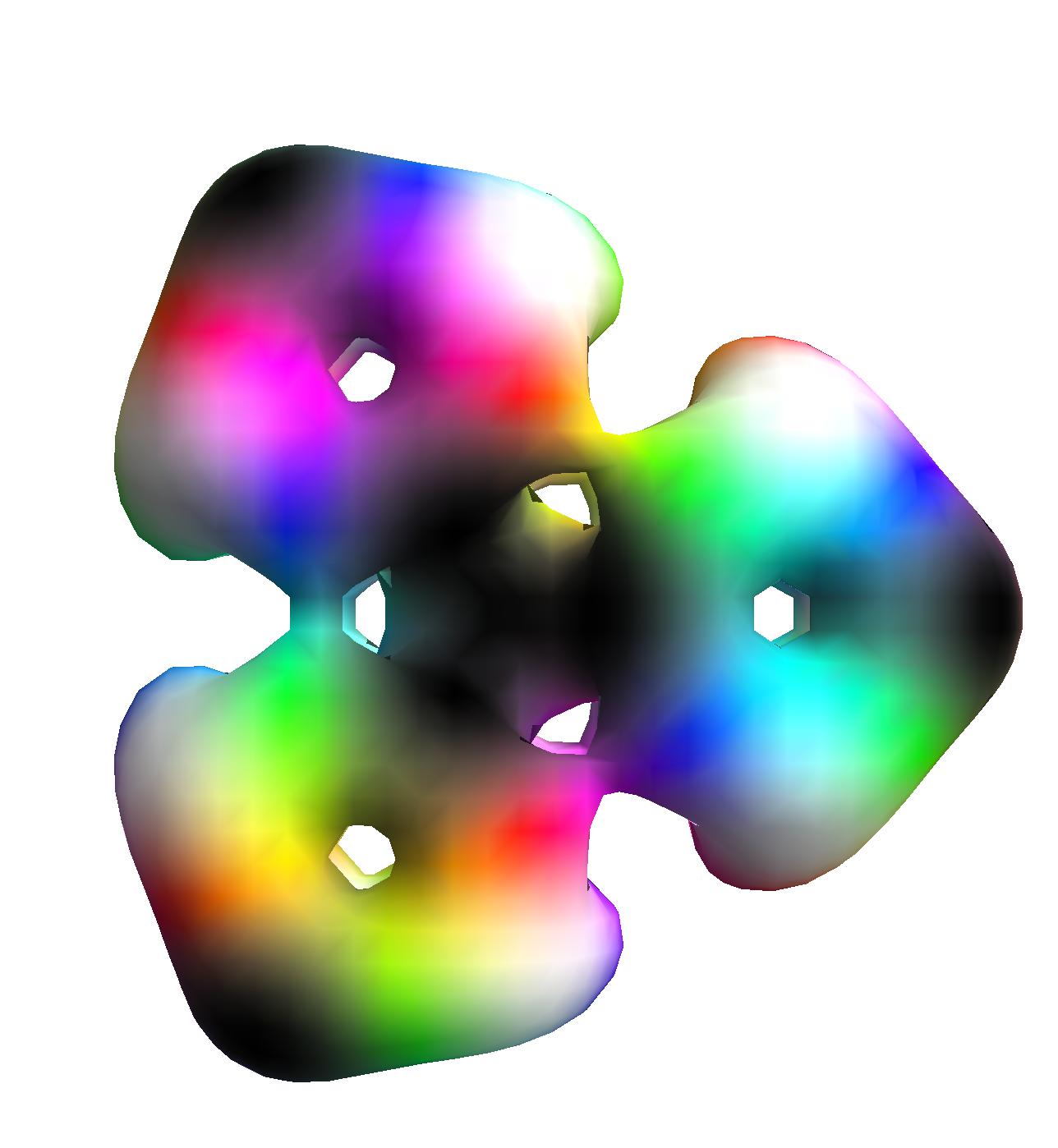}}
\subfigure[]{\includegraphics[totalheight=4.cm]{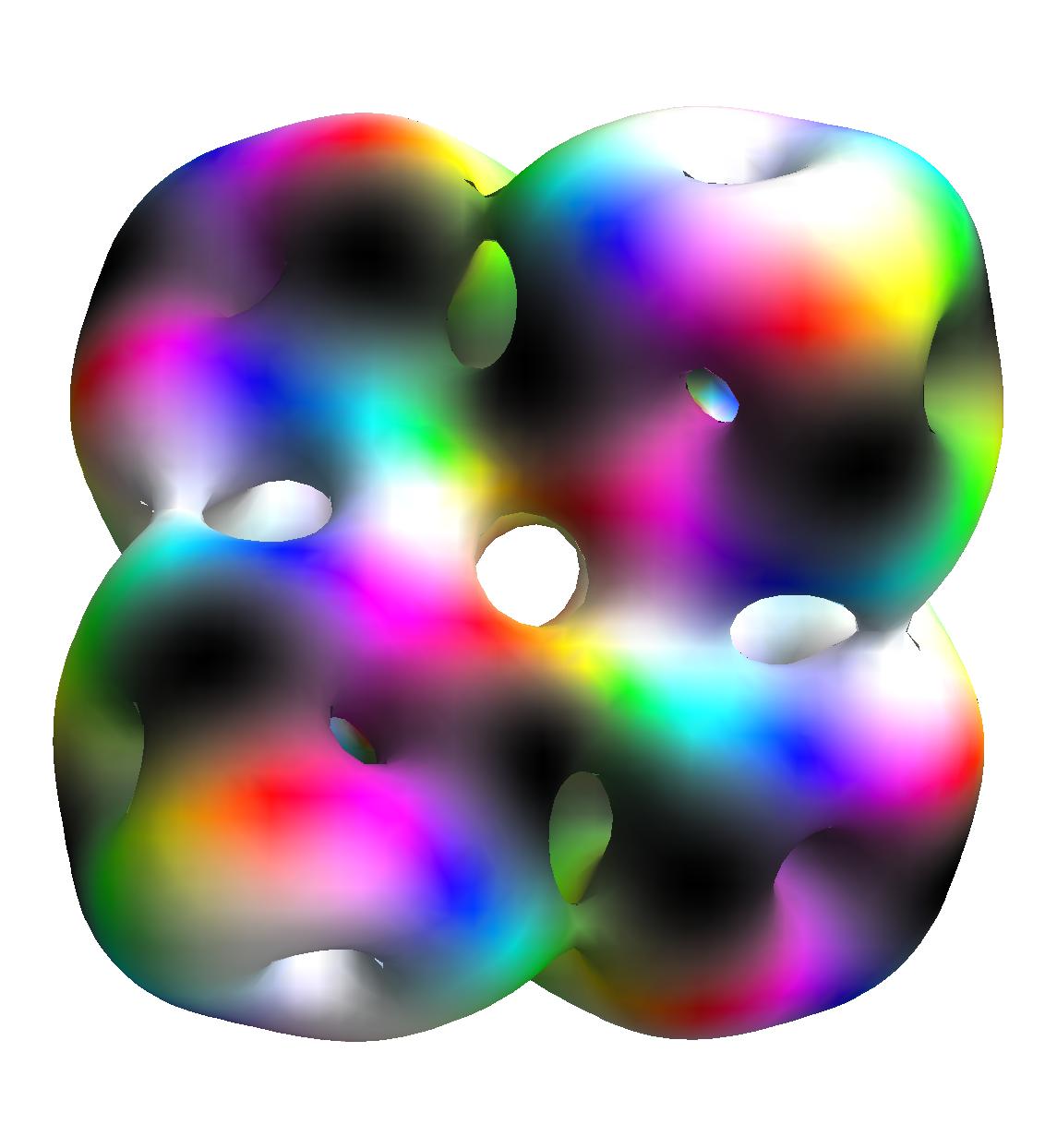}}
\subfigure[]{\includegraphics[trim = 1mm 1mm 1mm 1mm, clip, totalheight=4.cm]{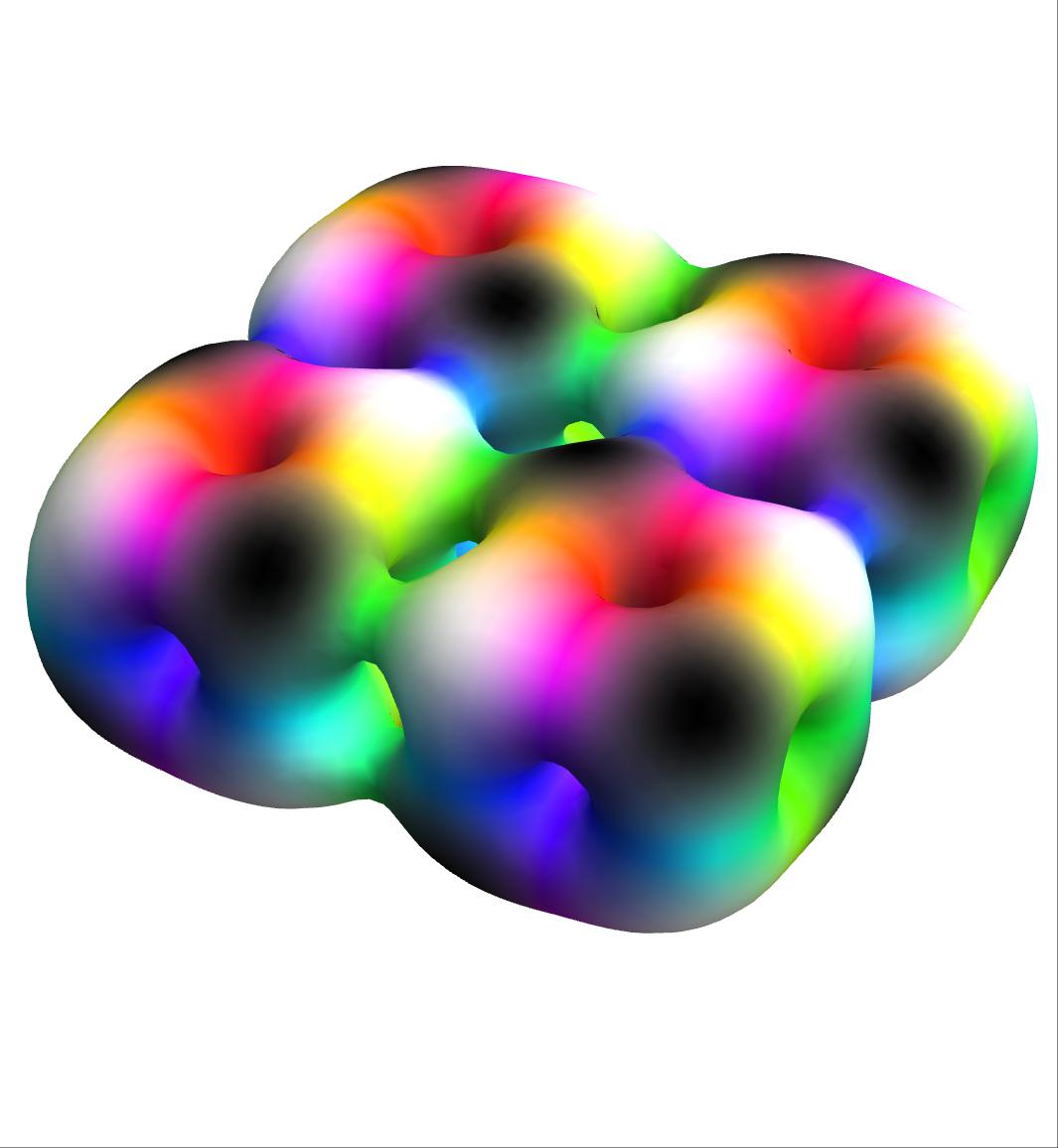}}\\
\subfigure[]{\includegraphics[trim = 1mm 3mm 1mm 1mm, clip, totalheight=5.cm]{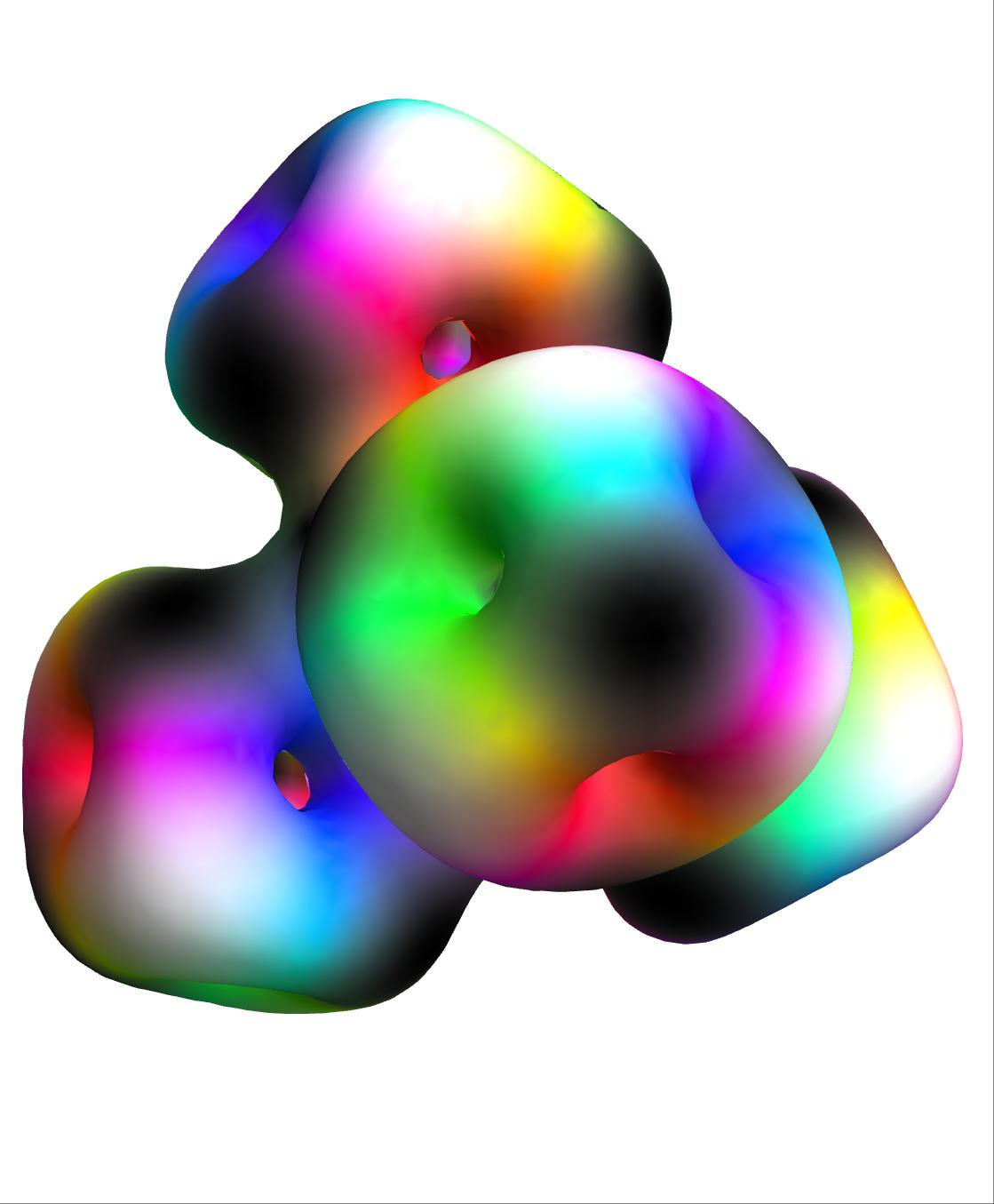}}
\subfigure[]{\includegraphics[totalheight=4.cm]{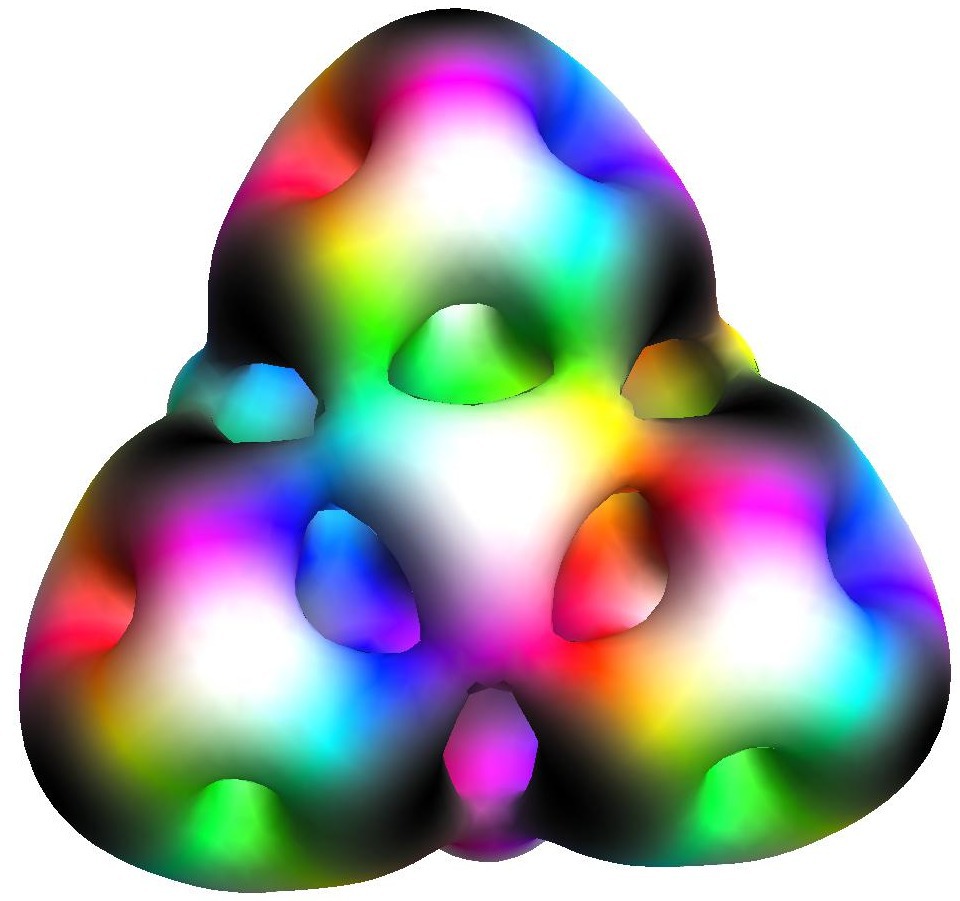}}
\subfigure[]{\includegraphics[totalheight=5.cm]{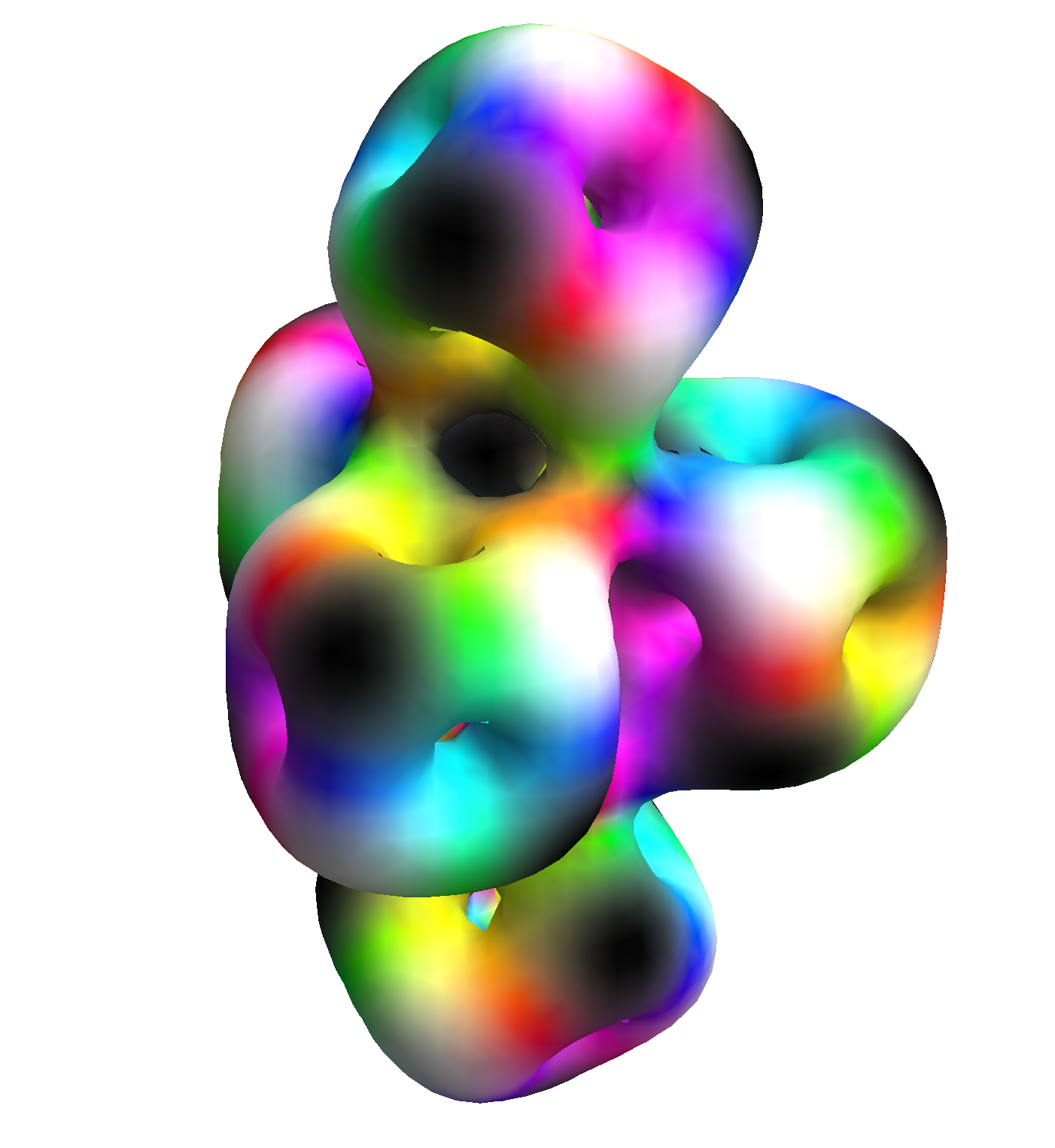}}\\
\subfigure[]{\includegraphics[trim = 1mm 1mm 1mm 1mm, clip, totalheight=4.cm]{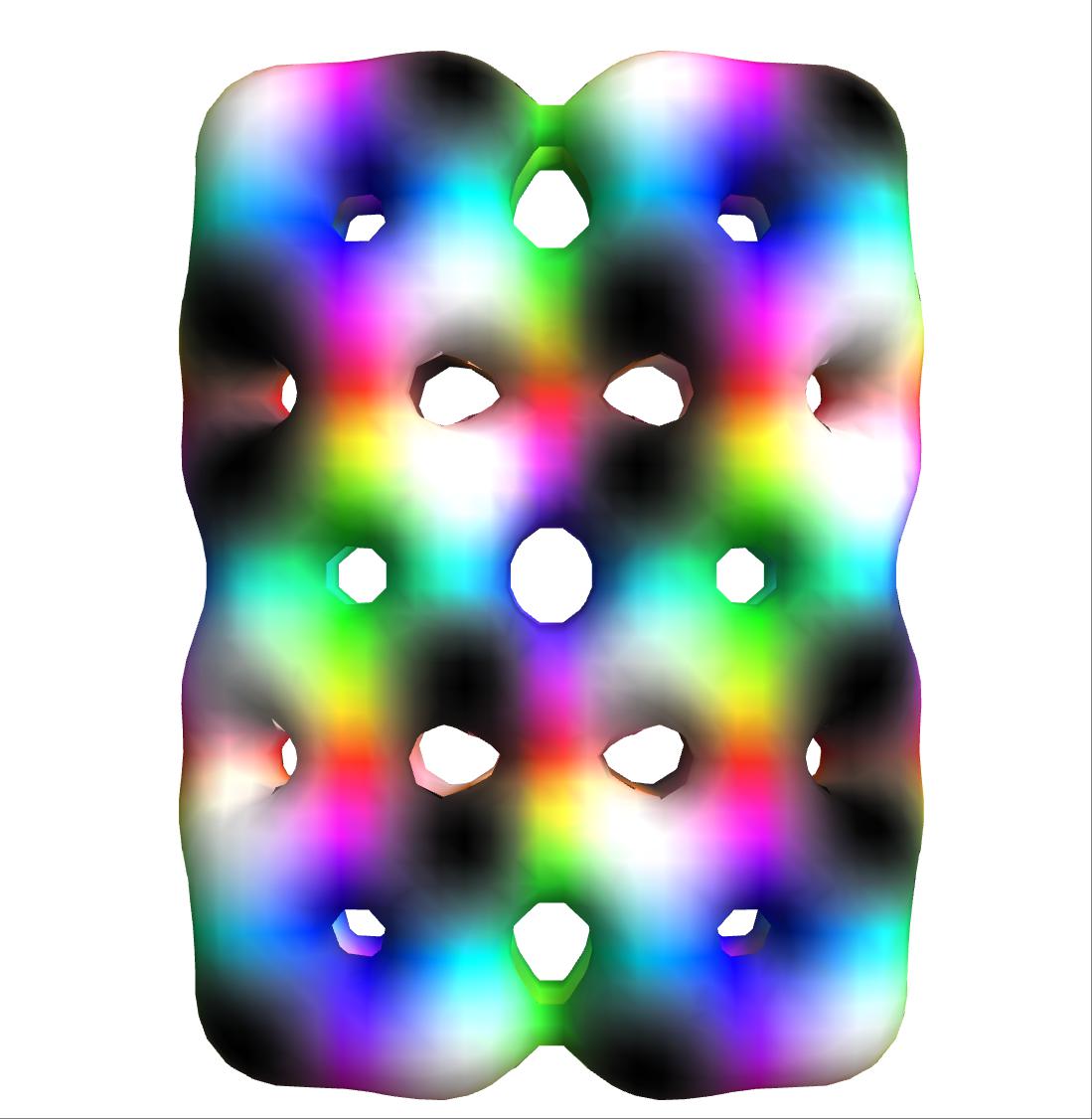}}
\subfigure[]{\includegraphics[totalheight=4.cm]{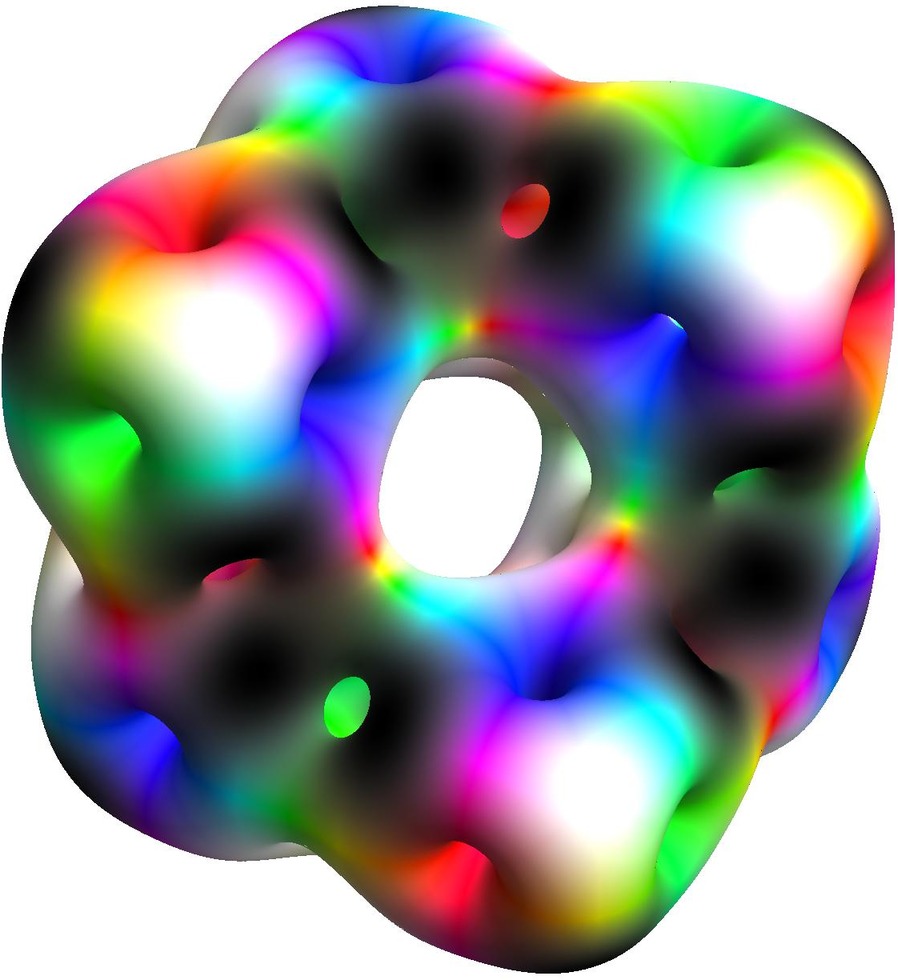}}
\subfigure[]{\includegraphics[trim = 1mm 1mm 1mm 1mm, clip, totalheight=4.cm]{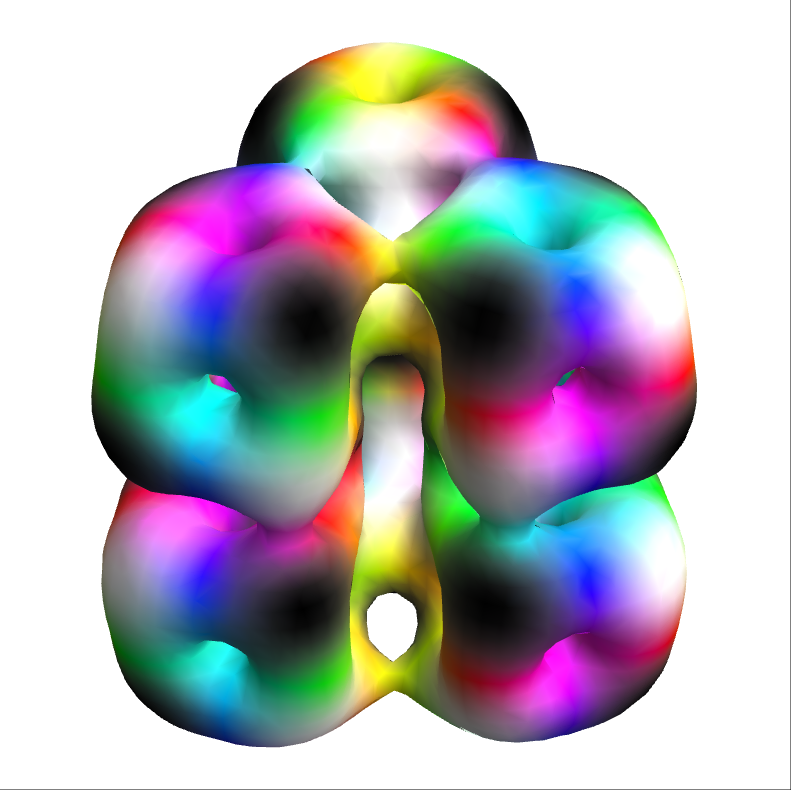}}
\subfigure[]{\includegraphics[totalheight=4.cm]{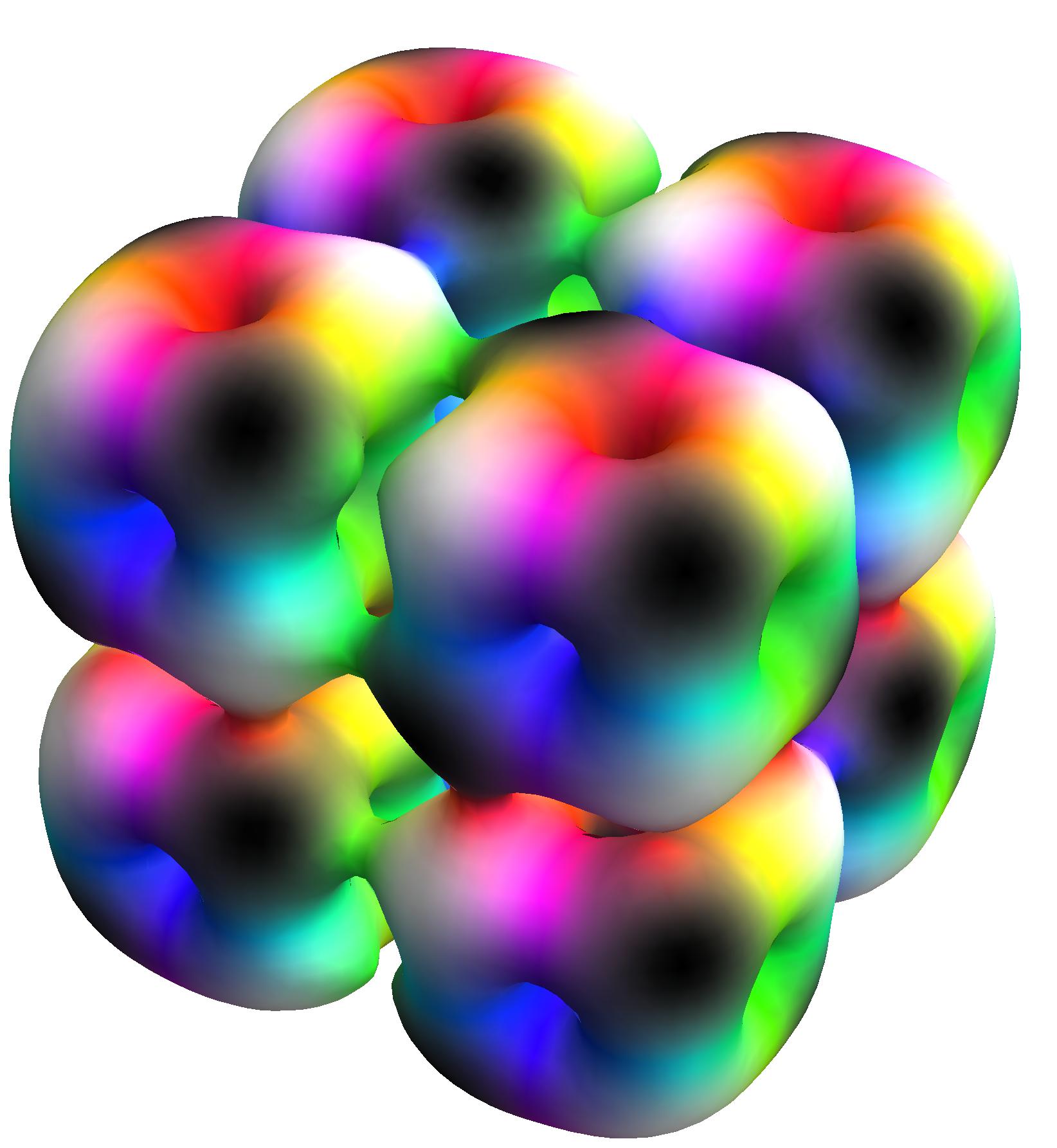}}
\caption{Surfaces of constant baryon density (not to scale) of Skyrmions with pion mass parameter $m=1$. The Skyrmions have baryon number and symmetry group: $B=8$ (a) $D_{4h}$ with $90^\circ$ twist, (b) $D_{4h}$ with no twist; $ \ B=12$ (c) $D_{4h}$, (d) $D_{3h}$; $ \ B=16$ (e) $D_{2d}$ (bent square), (f) $D_{4h}$ (flat square), (g) $T_d$; $ \ B=20$ (h) $T_{d}$, (i) $D_{3h}$; $ \ B=24$ (j) $D_{2h}$, (k) $D_{3d}$, (l) $D_{3h}$; $ \ B=32$ (m) $O_{h}$.}
\label{Sky_bary_dens}
\end{figure}

To find solutions, Skyrme fields of positive topological charge $B$ and with a given symmetry group $G$ are created by multi-layer rational map ans\"atze \cite{Feist:2012ps}, or product ans\"atze \cite{Battye:2006na}. These initial Skyrme field configurations are relaxed on grids with $(201)^3$ grid points and a spatial grid spacing $\Delta x=0.1$ to find precise solutions. We list in Table~\ref{Tab_Sky_inertia} the symmetry group, the energy and the diagonal elements of the isospin ($U_{ij}$), spin ($V_{ij}$) and mixed ($W_{ij}$) inertia tensors for Skyrmions with baryon numbers $B=8,\,12,\,16,\,20,\,24,\,32$.  The Skyrmions are orientated such that all off-diagonal elements of the inertia tensors vanish. 
The formulae for the inertia tensors $U_{ij},V_{ij},W_{ij}$ are rather complicated and have been given first in general form for arbitrary $S\!U(2)$ skyrmions in Refs.~\cite{Braaten:1988cc,Kopeliovich:1988np}. However, for numerical calculations it is much more convenient to express the inertia tensors in terms of the sigma field and pion field isotriplet, see formulae given in Refs.~\cite{Lau:2014sva,Battye:2014qva}. The baryon density isosurfaces we obtain are shown in Fig.~\ref{Sky_bary_dens}. On these surfaces, the $\boldsymbol{\pi}$-field values are visualized using Manton and Sutcliffe's field colouring scheme described in detail in Ref.~\cite{Manton:2011mi}.

In this paper, we calibrate the Skyrme model with properties of the Carbon-12 nucleus. The root mean square matter radius of a nucleus can be calculated within the Skyrme model as
\begin{align}
\left< r^2 \right>^{\frac{1}{2}}=\left(\frac{\int{r^2\,\mathcal{E}(\boldsymbol{x})\,\text{d}^3x}}{\int \mathcal{E}(\boldsymbol{x})\,\text{d}^3x}\right)^{\frac{1}{2}}\,,
\label{Sky_matrad}
\end{align}
where $\mathcal{E}(\boldsymbol{x})$ is the static energy density and $r=|\boldsymbol{x}|$. We list the matter radii $\left< r^2\right>^{\frac{1}{2}}_{\text{Sky}}$ (in Skyrme length units) of all the Skyrmions considered here in Table~\ref{Tab_en_rad}. The energy and length conversion factors are tuned to match the experimental nuclear mass 11178 MeV and matter radius $2.43$ fermi \cite{Tanihata:1986kh} for the Carbon-12 ground state. This fixes the conversion factors to be
\begin{align}
\left[\frac{F_\pi}{4e_{\text{Sky}}}\right]=6.154\,\text{MeV}\,, \quad 
\left[\frac{2}{e_{\text{Sky}}F_\pi}\right]=1.061\,\text{fm}\,,
\label{Cali_LM}
\end{align}
and gives the parameter values
\begin{align}
e_{\text{Sky}}=3.889,\quad F_\pi=95.6 \,\text{MeV},\quad \hbar=30.2\quad \text{ and } \quad m_\pi= 185.9 \,\text{MeV}\,, 
\label{Cali_LM_para}
\end{align}
where $\hbar=2e_{\text{Sky}}^2$. (More precisely, the energy unit is
$F_\pi / 4e_{\text{Sky}} = 6.154\,\text{MeV}$ and the length unit is
$2\hbar/e_{\text{Sky}}F_\pi = 1.061\,\text{fm}$, so the energy-length
unit is $\hbar/2e_{\text{Sky}}^2 = 6.529 \, \text{MeV}\,\text{fm}$. As
$\hbar = 197.3 \, \text{MeV}\,\text{fm}$ experimentally, $2e_{\text{Sky}}^2 = 30.2$ or equivalently $\hbar = 30.2$ in Skyrme units.) Note that we use in this article a value of $F_\pi$ that is substantially lower than its experimental value and a value of $m_\pi$ that is substantially larger than the physical pion mass. The experimental values for the pion mass and pion decay constant are given by $m_\pi=138\, \text{MeV}$ and $F_\pi=186\,\text{MeV}$, respectively. One can argue that $m_\pi$ and $F_\pi$ should be taken from experiment and that the Skyrme constant $e_{\text{Sky}}$ should be used as the fitting parameter.  Fitting Skyrmions with $m_\pi$ , $F_\pi$ and $F_K$ (with the kaon mass term added) at its physical values and $e_{\text{Sky}}\simeq4.1$ allows to reasonably describe the mass splittings \cite{Schwesinger:1991fy,Weigel:1995cz} within the $S\!U(3)$ multiplets of baryons. For these parameter values the absolute values of nuclear masses are not reproduced and one has to appeal to large Casimir energies from field fluctuations \cite{Moussallam:1993jd,Meier:1996ng} to make up the difference. Even for the $B=1$ Skyrmion, these quantum corrections to the soliton mass are difficult to calculate accurately and estimates have been given by various authors  \cite{Moussallam:1991rj,Moussallam:1993jd,Holzwarth:1995bv,Weigel:1994gi,Meier:1996ng,Scholtz:1993jg}.  Except for the work on Casimir energies of strongly bound $B=2$ configurations reported in Ref.~\cite{Scholtz:1993jg}, there do not exist estimates of Casimir effects for $B>1$. It is very difficult to pin down the exact magnitude of the Casimir contributions to Skyrmion masses because first of all this requires a full knowledge of the vibrational space of Skyrmions. New insights into the structure of vibrational spaces and their quantization have been reported recently in Ref.~ \cite{Halcrow:2015rvz}. In this article, we adopt a different point of view which goes back to Adkins, Nappi and Witten \cite{Adkins:1983ya}. It has been found effective when modelling nuclei by Skyrmions  to adjust the Skyrme parameters to fit nuclear masses and to interpret $m_\pi$  and $F_\pi$ as renormalized quantities \cite{Battye:2005nx,Houghton:2005iu,Fortier:2008yj}. For example, calibrating the Skyrme model with properties of the Carbon-12 nucleus has previously proven successful \cite{Lau:2014baa} in describing the spectrum of rotational excitations of Carbon-12, including the excitations of the Hoyle state, and in modelling nucleon-nucleon scattering within the Skyrme model \cite{Foster:2014vca,Foster:2015cpa}. In the following, we will refer to (\ref{Cali_LM}) as the Lau-Manton (LM) calibration. We will find that the LM calibration is well suited for calculating electromagnetic transition strengths within the Skyrme model. In addition, for this calibration the predicted nuclear masses and matter radii are in reasonable agreement with experimental data for a range of baryon numbers, see Table~\ref{Tab_en_rad}.

In the following, we will calculate and discuss the electric quadrupole transitions within the Skyrme model. Our discussion will be mainly focussed on the quadrupole transition strength, $B(E2)\hspace{-0.1cm}\uparrow$, between the $0^+$ ground state and the first $2^+$ state in even-even nuclei of zero isospin such as Carbon-12. The large quadrupole strength in the $0^+$ to $2^+$ transition in Beryllium-12 will serve as an example how transition strengths can be calculated in the presence of nonzero isospin.

\begin{table}[htb]
\caption{Skyrmions of baryon numbers $B=8,\,12,\, 16,\,20,\, 24,\,32$, for $m=1$. We list the symmetry group $G$ of each Skyrmion, its energy relative to the Skyrme-Faddeev bound $\frac{E}{12\pi^2B}$, and the diagonal elements of the inertia tensors $U_{ij},V_{ij},W_{ij}$ (in Skyrme units). We also list the isospin-zero 
  nuclei (in their ground states) that can be modelled by the Skyrmions. These are recognized from the symmetry group rather than the energy $E$.}
\begin{tabularx}{\textwidth}{clrrrrrrrrrrr}
\hline\hline
\ \quad $B$ \quad \qquad & $G$  & Nucleus & \quad $\frac{E}{12\pi^2B}$ & \ $U_{11}$ & \ $U_{22}$ & \ $U_{33}$ & \ $V_{11}$ & \ $V_{22}$ & \ $V_{33}$ & \ \ $W_{11}$ & \ \ $W_{22}$ & \ \ $W_{33}$\\
\hline
8 & $D_{4h}$ (twist) &\ce{^{8}_{4} Be_4}  & 1.279 & 298 & 292      & 326     & 4093 & 4094& 1381& 0 & 0&0\\
& $D_{4h}$ (no twist)  & &1.283 &287 & 291 & 350&  4615 &4615 & 1296& 0 &0 &0  \\
12 & $D_{4h}$ &Hoyle& 1.274 &440  & 449& 456 & 12137 & 12137& 2139 & 0& 0 &0  \\
     & $D_{3h}$           &\ce{^{12}_{6} C_6}& 1.278  & 442 & 442& 497& 5009& 5006& 7627& 41&   41& 38\\
16 & $D_{2d}$           (bent square)   & & 1.271& 572&571  & 674& 9123 & 9119& 14602& 0& 0 &0\\
 & $D_{4h}$           (flat square) & &1.272&   563& 567& 689& 9143 & 9174 & 15682& 0 & 0  &0\\
      &   $T_d$   &\ce{^{16}_{8} O_8} &1.276 &  586 &586&674&9100& 9101&  9128&0&0&0 \\ 
20&$T_d$     & & 1.273 &  757  & 757 & 819&  12820& 12820 &  12821& 0 & 0 &0\\
   & $D_{3h}$ &\ce{^{20}_{10} Ne_{10}}  &1.276 & 857 &   735 & 735 &18542  &18591   & 9762   & 15& -15 & -11\\
24  & $D_{2h}$ & & 1.267& 877 & 862 & 956 & 26980 & 14189 & 36783 & 0& 0 &0 \\
 & $D_{3d}$         &&1.269&  879  & 890  & 959 &19600   & 19600 &  29863 &0  & 0 &0  \\
  & $D_{3h}$  &\ce{^{24}_{12} Mg_{12}}         &1.269&  869    & 869  & 1006 &20554 & 20454 & 16226   & -99 & 99 &99  \\
32&$O_h$      & \ce{^{32}_{16} S_{16}}     & 1.264 & 1115   & 1116  & 1367 &  31625 &31628   & 31704 & 0  &0 &0 \\\hline\hline
\end{tabularx}
\label{Tab_Sky_inertia}
\end{table}

\begin{table}[htb]
\caption{Nuclear masses $E$ and root mean square matter radii $\left<r^2 \right>^{\frac{1}{2}}$ for nuclei of baryon numbers $B=8,\,12,\, 16,\,20,\, 24,\,32$ and isospin zero. Here, the subscripts ``Sky'' and ``LM'' refer to Skyrme units and the Lau-Manton calibration (\ref{Cali_LM}), respectively. The experimental matter radii $\left< r^2 \right>^{\frac{1}{2}}_{\text{Exp}}$ are taken from Ref.~\cite{2013ADNDT..99...69A} and are given in fermi. Note that due to its instability there are no data available for Beryllium-8, so here (${}^\ast$) we give the charge radius for its isobar  \ce{^{8}_{3} Li_5}. The Hoyle state's nuclear radius (${}^{\ast\ast}$) has been measured in Ref.~\cite{Danilov:2009zz}. The experimental matter radius for the Carbon-12 ground state (${}^{\ast\ast\ast}$) is taken from Ref.~\cite{Tanihata:1986kh}. }
\begin{tabularx}{\textwidth}{clrrrrrrl}
\hline\hline
\ \ \ $B$ \ \ \qquad & $G$ & Nucleus & \quad $E_{\text{LM}}$ [MeV] & \ \ $E_{\text{Exp}}$ [MeV]& \ \ $\left< r^2 \right>^{\frac{1}{2}}_{\text{Sky}}$  & \ \ $\left< r^2 \right>^{\frac{1}{2}}_{\text{LM}}$ $[\text{fm}]$   & \ \ $\left< r^2 \right>^{\frac{1}{2}}_{\text{Exp}}$ $[\text{fm}]$ & \\
\hline
8  &$D_{4h}$ (twist)      &\ce{^{8}_{4} Be_4} &  7457.5&7451.9&2.05 &2.18&   2.34 & ${}^{\ast}$  \\
  &$D_{4h}$ (no twist)  &&7480.9&&2.15 &2.28& & \\
 12    &   $D_{4h}$             & Hoyle & 11142.6 &&2.76 &2.93& 2.89 & $^{\ast\ast}$\\
        &$D_{3h}$                &  \ce{^{12}_{6} C_6} &11178&11178&2.29  & 2.43& 2.43 &$^{\ast\ast\ast}$  \\ 
16 &     $D_{2d}$    (bent square)  & &14821.9&&2.66 &2.82& & \\
 &     $D_{4h}$     (flat square)  & &14833.5& &2.70&2.87& &  \\
 &     $T_{d}$     &\ce{^{16}_{8} O_8} &14903.5&14903.9&2.35 &2.50&   2.70 & \\
20 &    $T_d$    & &  18556.5& & 2.60         &2.76&  & \\
     &     $D_{3h}$          &\ce{^{20}_{10} Ne_{10}}  & 18600.2  &    18629.8      &  2.86  &     3.03        & 3.01 &\\
 24 &    $D_{2h}$          &    &22170.1&&3.32 & 3.53& & \\
  &    $D_{3d}$          &  &22197.8& &3.13& 3.33&  & \\
 &  $D_{3h}$  &\ce{^{24}_{12} Mg_{12}}  &22212.1&22355.8&2.87&3.04&3.06   & \\
32 &    $O_h$           &\ce{^{32}_{16} S_{16}} &29480.5&  29807.8&    3.17            &3.36&3.26 & \\\hline\hline
\end{tabularx}
\label{Tab_en_rad}
\end{table}

\section{Electromagnetic Transition Strengths in the Skyrme Model}

In the Skyrme model, the electric charge density $\rho(\boldsymbol{x})$ \cite{Witten1983422,Witten1983433} is given by
\begin{eqnarray}
  \rho(\boldsymbol{x})=\frac{1}{2}\mathcal{B} (\boldsymbol{x})+ \mathcal{I}_3(\boldsymbol{x})\,,
\label{Sky_current}
\end{eqnarray}
where $\mathcal{B}(x)$ denotes the baryon density, the integrand of (\ref{Sky_bary}), and $\mathcal{I}_3(x)$ is the third component of the isospin density. For quantum states with zero isospin, the charge density $\rho$ is half the baryon density \cite{Manton:2006tq}, and the total electric charge is $\frac{1}{2}B$ (in units of the proton charge $e$). Nuclei with zero isospin have equal numbers of protons and neutrons. For nonzero isospin, the isospin density $\mathcal{I}_3(\boldsymbol{x})=\omega\mathcal{U}_{33}(\boldsymbol{x})$ contributes to the total electric charge. Here, $\omega$ is the isorotational angular frequency and the isospin inertia density is \cite{Lau:2014sva,Battye:2014qva}
\begin{align}
\mathcal{U}_{ij}(\boldsymbol{x})=2\Bigg\{\left(\boldsymbol{\pi}\cdot\boldsymbol{\pi}\delta_{ij}-\pi_i\pi_j\right)\left(1+\partial_k\sigma\partial_k\sigma+\partial_k\boldsymbol{\pi}\cdot\partial_k\boldsymbol{\pi}\right)-\epsilon_{ide}\epsilon_{jfg}\left(\pi^d \partial_k\pi^e\right)\left(\pi^f\partial_k\pi^g\right)\Bigg\}\,\,.\label{U_inertia}
\end{align}

The classical electric quadrupole tensor for a Skyrmion is defined as 
\begin{eqnarray}
Q_{ij}=\int \text{d}^3 x\,\left(3x_ix_j-|\boldsymbol{x}|^2\delta_{ij}\right)\rho(\boldsymbol{x})\,.
\label{Sky_quadrupole}
\end{eqnarray}
For a Skyrmion in its standard orientation, the tensor is diagonal and
the diagonal entries $Q_{11}, Q_{22}, Q_{33}$ are the quadrupole
moments. The quadrupole tensor is traceless (up to numerical
inaccuracies) so $Q_{11} + Q_{22} + Q_{33} = 0$. Almost all the
Skyrmions we consider here can be orientated so that they have a 
cyclic symmetry greater than $C_2$ along the 3-axis. Then 
$Q_{11} =Q_{22}$ and $Q_{33}$ is the quadrupole moment of largest magnitude.

\subsection{Zero Isospin}

We list in Table~\ref{Tab_Q} our numerical results for quadrupole moments of Skyrmions and the corresponding nuclei with mass numbers $B=8,\,12,\,16,\,20,\,24,\,32$ and zero isospin. Here, the charge density $\rho$ is half the baryon density. The Skyrme model's predictions are given in Skyrme units and can be converted to physical units by multiplying by the square of the length scale. We orientate the classical Skyrmion such that $Q_{33}$ is the quadrupole moment of maximal magnitude, as discussed above. Then, the nucleus' intrinsic electric quadrupole moment  $Q_{0}$ can be identified with $Q_{33}\times\left[\frac{2}{e_{\text{Sky}}F_\pi}\right]^2$. With the calibration (\ref{Cali_LM}), we obtain the intrinsic quadrupole moments $Q_0$ (in units of electron barn, $\text{e}\text{b}$) which are given in the penultimate column of Table~\ref{Tab_Q}.  For comparison, we also list experimental data, where available. Recall that a factor of $\frac{1}{100}$ is required to convert $[\text{fm}]^2$ to barn.

\begin{table}[htb]

\centering

\caption{Intrinsic quadrupole moments for Skyrmions of mass numbers $B=8,\,12,\, 16,\,20,\, 24,\,32$ and of zero isospin. ``Sky'' and ``LM'' refer to Skyrme units and the Lau-Manton calibration (\ref{Cali_LM}), respectively. We use ``---'' to denote Skyrmions of zero quadrupole moment. Unless otherwise stated, the experimental results for intrinsic quadrupole moments (in electron barn) are taken from Ref.~\cite{Raman:1201zz} and have been derived from experimental $B(E2)$ transition strengths via Eq.~(\ref{E2transit}). Note that we state two different intrinsic quadrupole moments for the Beryllium-8 nucleus. These values are obtained by   (${}^\text{I}$) a variational Monte Carlo (VMC) calculation \cite{Wiringa2000} and  (${}^{\text{II}}$) a Green's function Monte Carlo (GFMC) method \cite{PhysRevLett.111.062502}. The experimental quadrupole moment given for Oxygen-16 \cite{Raman:1201zz} is bracketed since this value has been derived from experimental $B(E2)$ transition strengths from the $0_1^+$ ground state to the first-excited $2^+$ state. Within the Skyrme model description, this $E2$ transition corresponds to an inter-band transition and hence cannot be modelled using the techniques described in this paper. See discussion and Fig.~\ref{B16_rotband} in subsequent section on rotational states and transitions in Oxygen-16 for more details.}
\begin{tabularx}{\textwidth}{clrrrrrr}
\hline\hline
\qquad \ $B$ \qquad \qquad &$G$& Nucleus & \quad $Q_{11}^{\text{Sky}}$  & \quad $Q_{22}^{\text{Sky}}$ & \quad $Q_{33}^{\text{Sky}}$  & \quad $Q_{0}^{\text{LM}}$ $[\text{e}\text{b}]$ & \qquad \qquad $Q_{0}^{\text{Exp}}$ $[\text{e}\text{b}]$\\
\hline
8 &    $D_{4h}$ (twist) & \ce{^{8}_{4} Be_4}        &  -8.54& -8.55 &17.10& +0.192& $+0.266^{\text{I}},\,+0.320^{\text{II}}$ \\
   &    $D_{4h}$ (no twist)            &                                    &-10.5&  -10.5&  21.1 & +0.238&  \\
 12 &  $D_{4h}$      &Hoyle                       &-32.1&-32.1&   64.3& +0.724 &\\
 &  $D_{3h}$  &  \ce{^{12}_{6} C_6}  &8.99&9.10& -18.0  &-0.203  &-0.200 \\ 
16 &  $D_{2d}$  (bent square) &     &18.2&18.4& -36.6 &  -0.412  &(0.202)\\
 &  $D_{4h}$  (flat square) &   &21.2&21.3& -42.5& -0.478&\\
 &  $T_{d}$  &\ce{^{16}_{8} O_8}   &--- & --- & --- & ---   & --- \\
 20 &  $T_{d}$  & &---&---&---&---& \\
 & $D_{3h}$  &\ce{^{20}_{10} Ne_{10}} & -18.8 &-18.6  & 37.4 & +0.421&+0.584\\
 24  & $D_{2h}$    & &-107&116 &-9.16 & -0.103  & \\
       &  $D_{3d}$  &   &34.4&34.4&-68.9& -0.776 &  \\
  &  $D_{3h}$   &\ce{^{24}_{12} Mg_{12}} & -13.6& -12.5&26.1&+0.294 &  +0.659\\
32 &  $O_{h}$  &  \ce{^{32}_{16} S_{16}} &--- &---&   ---  &---& +0.549 \\\hline\hline
\end{tabularx}
\label{Tab_Q}
\end{table}

For a nuclear state, let $J$ be the total angular momentum and $k$ its projection on the body-fixed 3-axis. The reduced electric quadrupole transition strength $B(E2)$ from an initial state $| J_i \,, k \rangle$ to a final state $| J_f \,, k \rangle$ can be obtained \cite{bohrMot1,bohr1998nuclear2} from the intrinsic moment $Q_0$ via
\begin{align}
B(E2: J_i \,, k \rightarrow J_f \,, k)=\frac{5}{16\pi} Q_0^{2}\,\left\langle J_i\,k ; 2\,0 \big| J_f\,k \right\rangle^2\,,
\label{BE2_eq}
\end{align}
where the Clebsch-Gordan coefficient $\left\langle J_i\, k;2\,0 \big| J_f\,k\right\rangle$ governs the coupling of the angular momenta.  

For electromagnetic transitions between states  $J_i=J$ and $J_f=J+2$, with $k=0$, the Clebsch-Gordan coefficient in (\ref{BE2_eq}) simplifies to 
\begin{eqnarray}
\left\langle J\,0;2\,0 | (J+2)\,0\right\rangle^2=\frac{3(J+1)(J+2)}{2(2J+1)(2J+3)}\,.
\end{eqnarray}
Hence, the reduced electric quadrupole transition probability, $B(E2)\hspace{-0.1cm}\uparrow$, from the spin $0^+$ ground state to the first excited spin $2^+$ state is given by 
\begin{eqnarray}\label{E2transit}
B(E2:0^+\rightarrow 2^+)=\frac{5}{16\pi} Q_0^2\,.
\end{eqnarray}

Note that electromagnetic excitation $B(E2)\hspace{-0.1cm}\uparrow$ and decay $B(E2)\hspace{-0.1cm}\downarrow$ of a nuclear state are related \cite{bohrMot1,bohr1998nuclear2} by
\begin{eqnarray}
B(E2:J_f\rightarrow J_i)=\frac{2J_i+1}{2J_f+1}B(E2:J_i\rightarrow J_f)\,.
\label{Trans_up}
\end{eqnarray}

By substituting the intrinsic quadrupole moments $Q_{0}^{\text{LM}}$ listed in Table~\ref{Tab_Q} in Eq.~(\ref{E2transit}), we obtain the Skyrme model's predictions for the $B(E2)\hspace{-0.2cm}\uparrow$ values  (in units of $\text{e}^2\text{b}^2$) for nuclei of mass numbers $B=8,\,12,\, 16,\,20,\, 24,\,32$; they are presented in the fifth column of Table~\ref{Tab_BE2}, with experimental data in the sixth column. In the fourth column of Table~\ref{Tab_BE2}, we list the corresponding $B(E2)\hspace{-0.2cm}\uparrow$ values in Skyrme units. These $B(E2)^{\text{Sky}}$ values are obtained by substituting $Q_{33}^{\text{Sky}}$ given in Table~\ref{Tab_Q}   in Eq.~(\ref{E2transit}). They are related to physical units by the factor $\left[\frac{2}{e_{\text{Sky}}F_\pi}\right]^4$, the fourth power of the Skyrme length unit. We also include in Table~\ref{Tab_BE2} the calculated transition strength $B(E2:0^+\rightarrow 2^+)$ for the short-lived \ce{^{12}_{4} Be_8} nucleus, to be discussed below, and for the Hoyle state of \ce{^{12}_{6} C_6}.  

To further simplify comparison with experimental data we convert to Weisskopf units W. This compares the transition strength with the single-particle strength
\begin{align}
B(E2)\hspace{-0.1cm}\uparrow_{\text{sp}}=2.97\times 10^{-5}B^{\frac{4}{3}}\ \text{e}^2\text{b}^2\,.
\label{Weisskopf}
\end{align}
The strength in Weisskopf units is $W = B(E2)\hspace{-0.1cm}\uparrow \, / \, B(E2)\hspace{-0.1cm}\uparrow_{\text{sp}}$, and is a measure of collective quadrupole effects in nuclei. A value higher than 5 indicates substantial collectivity. 

In the following subsections, we discuss each nucleus separately. The structure and excitation spectrum of Oxygen-16 are particular difficult to understand within a shell-model description \cite{Nagai01041962}. Recent progress has been made via \emph{ab initio} calculations using alpha cluster initial states with tetrahedral and square configurations \cite{Epelbaum:2013paa}. Within the Skyrme model, tetrahedral and square-like configurations of charge-4 sub-units arise as $B=16$ Skyrmion solutions \cite{Battye:2006na}. For this reason, we devote a separate, longer section of this paper to rotational states and transitions in Oxygen-16. 

\subsubsection{Beryllium-8}

For Beryllium-8, we calculate $B(E2:0^{+}\rightarrow 2^{+})$ transition strengths using the two known $D_{4h}$-symmetric Skyrmions. For the twisted $B=8$ Skyrmion we find $B(E2)=  0.00366\, \text{e}^2\text{b}^2$ and for the untwisted Skyrmion $B(E2)= 0.00563 \, \text{e}^2\text{b}^2$. Due to the instability of Beryllium-8 to alpha decay, we are unable to compare our results with actual experimental data. Instead, we include in Table~\ref{Tab_BE2} $B(E2)$ values based on Hartree-Fock calculations \cite{Raman:1201zz} and on Monte Carlo methods \cite{Wiringa2000,PhysRevLett.111.062502}. Note that the available theoretical values vary significantly depending on which model is used. This makes it impossible to test the accuracy of our Skyrme model $B(E2)$ predictions. Our predicted intrinsic quadrupole moments are consistent with the prolate shape found in Hartree-Fock and Monte Carlo calculations.

\subsubsection{Carbon-12 and Hoyle State}

For $B=12$, rotational excitations of the $D_{3h}$ triangular Skyrmion solution match the Carbon-12 ground state band, and excitations of the $D_{4h}$ chain solution reproduce the Hoyle band \cite{Lau:2014baa}. The $D_{3h}$-symmetric Skyrmion has the oblate shape assumed for the Carbon-12 nucleus \cite{Esbensen:2011bg}. Our calculated intrinsic quadrupole moment $Q_0=-0.203\, \text{e}\text{b}$ agrees well with the experimental value $Q_0=-0.200\, \text{e}\text{b}$ \cite{Raman:1201zz} extracted from the measured strength of the $0_1^{+}\rightarrow2_1^{+}$ transition. The associated transition strength $B(E2:0_1^{+}\rightarrow2_1^{+})=0.00409\,\text{e}^2\text{b}^2$ deviates by $3\%$ from the experimental value.

Measuring the $E2$ transition strength from the $2_2^{+}$ Hoyle state
to the $0_2^{+}$ Hoyle state is experimentally challenging
\cite{Freer:2014qoa} and would require a highly efficient
particle-gamma experimental setup \cite{2014LNP...875...25J}. 
Interpreting the Hoyle state as a linear chain formed out of three 
$B=4$ Skyrmions, we predict the transition strength in the opposite
direction, $B(E2:0_2^{+}\rightarrow2_2^{+})=0.0521\,\text{e}^2\text{b}^2$. This
corresponds to $63.9$ W, arising from a strongly prolate intrinsic
shape with an intrinsic quadrupole moment $Q_0=0.724\,
\text{e}\text{b}$.

\begin{table}[htb]
\caption{Quadrupole transition strengths $B(E2)\hspace{-0.15cm}\uparrow$ for nuclei of baryon numbers $B=8,\,12,\, 16,\,20,\, 24,\,32$. ``Sky'' and ``LM'' refer to Skyrme units and the Lau-Manton calibration (5), respectively. ``---'' denotes zero transition strength. Unless otherwise stated, the experimental $B(E2)\hspace{-0.1cm}\uparrow$ values are taken from Ref.~\cite{Raman:1201zz}.  Note that we state three different estimated transition strengths for the Beryllium-8 nucleus. These are obtained by (${}^{\text{I}}$) Hartree-Fock+BCS calculations with the Skyrme SIII force \cite{Raman:1201zz}, (${}^{\text{II}}$) a variational Monte Carlo (VMC) calculation \cite{Wiringa2000} and  (${}^{\text{III}}$) a Green's function Monte Carlo (GFMC) method \cite{PhysRevLett.111.062502}. (${}^{\text{IV}}$) For the short-lived Beryllium-12 isotope we obtain the estimated experimental $B(E2)\hspace{-0.1cm}\uparrow$ value by multiplying by 5 the $B(E2)\hspace{-0.1cm}\downarrow$ value measured in Ref.~\cite{Imai:2009zza}.}
\begin{tabularx}{\textwidth}{clrrrlrlr}
\hline\hline
\ \ \ $B$ \ \ \qquad &        $G$     & Nucleus & $ \quad
B(E2)^{\text{Sky}}$   &$ \ B(E2)^{\text{LM}}$   &$[\text{e}^2\text{b}^2]$
&$B(E2)^{\text{Exp}}$ &$[\text{e}^2\text{b}^2]$     &Dev. $[\%]$ \\
\hline
8      &$D_{4h}$ (twist)       &\ce{^{8}_{4} Be_4}    &      29.1      &  0.00366 &  ( 7.7 W  )   &  $0.003$&$^{\text{I}}$  &22\%\\
      &     &     &     &    &     & $0.0100$&$^{\text{II}}$  &63.3\%\\
      &     &     &     &    &     & $0.00740$&$^{\text{III}}$  &50.5\%\\
8  &$D_{4h}$ (no twist)      &                            &      44.6                              &       0.00563  &   (11.8 W)     &          &   &\\


 12    &   $D_{4h}$                 &   Hoyle            &     411                          &0.0521 & (63.9 W)&&&\\
 &$D_{3h}$                    &\ce{^{12}_{6} C_6}       &32.5& 0.00409 & (5.0 W)& 0.00397 & (4.9 W) &3.02\%\\
     &$D_{3h}$                    &\ce{^{12}_{4} Be_8}      &  14.2 &  0.00181 & (2.2 W)&$0.0040$& (4.9 W)$^{\text{IV}}$ &54.7\%\\
16 &     $D_{2d}$           (bent square)         & \ce{^{16}_{8} O_8}  &133  &0.0168 & (14.1 W)  && &\\
     &     $D_{4h}$       (flat square)         &   & 179& 0.0227 & (18.9 W) && &\\
20     & $D_{3h}$                  &                    \ce{^{20}_{10} Ne_{10}}                                         & 139    & 0.0176 & (10.9 W) &0.0340 & (21 W) & 48.1\%\\
24 &      $D_{3h}$                                   &\ce{^{24}_{12} Mg_{12}}&  68.1   & 0.00864 & (4.2 W)&0.0432 & (21 W)&80.0\%\\
32 &    $O_h$                                         &\ce{^{32}_{16} S_{16}}   &     ---&---&&0.0300& (9.8 W)& \\  \hline\hline
\end{tabularx}
\label{Tab_BE2}
\end{table}

\subsubsection{Neon-20}

In the $\alpha$-particle model, Neon-20 is described in terms of five
$\alpha$-particles arranged in a triangular bipyramid
\cite{Wefelmeier1937}. Four of the five low-lying rotational bands in
Neon-20 can be understood \cite{Bouten1962} using this bipyramidal
$\alpha$-particle arrangement. In the Skyrme model, there exists an
analogous bipyramidal cluster arrangement \cite{Battye:2006na} of five
$B=4$ cubes (see Fig.~\ref{Sky_bary_dens} (i)). This
$D_{3h}$-symmetric configuration is not the global minimal energy
Skyrmion with $B=20$, but a nearby saddle point solution. In agreement
with experimental data, this bipyramidal Skyrmion structure gives a
prolate deformed Neon-20 ground state. The associated intrinsic
quadrupole moment $Q_0=0.421\, \text{e}\text{b}$ is less than the
experimental value $Q_0=0.584\, \text{e}\text{b}$ \cite{Raman:1201zz}
deduced from the measured $B(E2)\hspace{-0.1cm}\uparrow$ value. For
the electric quadrupole transition from the $0^{+}$ ground state to
the first excited $2^+$ state, the corresponding $B(E2)= 0.0176\,
\text{e}^2\text{b}^2$ is approximately $50\%$ less than the
experimental value. We have not yet identified any states of Neon-20
with the quantized states of the $T_d$-symmetric $B=20$ Skyrmion.

\subsubsection{Magnesium-24}

For $B=24$, we consider three very different Skyrmion solutions: a non-planar $D_{3d}$-symmetric ring formed of six $B=4$ Skyrmion cubes with each neighbouring pair being rotated through $90^\circ$ around the line joining the cubes (see baryon density isosurface in Fig.~\ref{Sky_bary_dens}~(k) and Ref.~\cite{Feist:2012ps}), a triaxial configuration constructed by gluing together two linear $B=12$ Skyrmions (see Fig.~\ref{Sky_bary_dens}~(j)), and two triangular $B=12$ Skyrmions bound together into a $B=24$ solution (see Fig.~\ref{Sky_bary_dens}~(l)). The ring was previously believed to be the Skyrmion of minimal energy, but at least one of the other, newly found solutions appears to have lower energy.

Among these Skyrmions, we find that Magnesium-24 is probably best described by the $D_{3h}$-symmetric solution made of two triangular $B=12$ Skyrmions. The quadrupole moment is found to be $Q_0= 0.294\,\text{e}\text{b}$ which is still significantly less than the experimental value $Q_0= 0.659\, \text{e}\text{b}$. The corresponding quadrupole transition strength $B(E2)= 0.00864  \, \text{e}^2\text{b}^2$ is much lower than the experimental value $B(E2)= 0.0432  \,\text{e}^2\text{b}^2$. The quadrupole moment of the Skyrmion ring solution has the wrong sign (compare Table~\ref{Tab_Q}) and does not reproduce the prolate ground state of Magnesium-24. The new $B=24$ solution in Fig.~\ref{Sky_bary_dens}~(j) is triaxial and is badly approximated as an axially symmetric solution. Hence our analysis cannot be applied to this Skyrmion solution. However, this might give a better quadrupole moment and $B(E2)$ value. 

\subsubsection{Sulphur-32}

Calculating nuclear properties of Sulphur-32 has proven to be difficult in the past \cite{Bauhoff:1980zz} and earlier work using Hartree-Fock calculations on the rotational spectra in Sulphur-32 yielded contradictory, model-dependent results. The experimental excitation energies of the  $0^+$, $2^+$ and $4^+$ states of Sulphur-32 agree very well with the vibrational excitations of a spherically shaped nucleus \cite{PhysRevC.58.699}. However, experimentally Sulphur-32 possesses a relatively large positive quadrupole moment \cite{Raman:1201zz,Stone200575} which suggests a significant prolate nuclear deformation. This can be understood within the nuclear coexistence model \cite{PhysRevC.9.1192} for Sulphur-32, in which spherical and prolate rotational bands coexist.

In the Skyrme model, Sulphur-32 is modelled by the cubically symmetric $B=32$ Skyrmion (see baryon density isosurface in Fig.~\ref{Sky_bary_dens} (m)  and Ref.~\cite{Battye:2006na}) and hence its intrinsic quadrupole moment vanishes. However, this Skyrmion is still of interest because it is a candidate to model the vibrational excitations of Sulphur-32. As the Skyrmion spins, the Skyrmion deforms \cite{Battye:2005nx,Houghton:2005iu,Fortier:2008yj,Battye:2014qva} and a non-zero quadrupole moment will be induced.  A calculation of $E2$ transitions for non-rigidly spinning Skyrmion solutions requires different techniques and is beyond the scope of this paper.

\section{Rotational States and Transitions in Oxygen-16}\label{Sec_O16}

Oxygen-16 has previously been investigated within the alpha cluster model \cite{Buck:1975zz} and by performing lattice effective field theory calculations \cite{Epelbaum:2013paa}. The results suggest that there exist two rotational bands, one based on a tetrahedral arrangement of alpha clusters, and another on a square-like arrangement. This section discusses whether such an interpretation is possible in the Skyrme model.

Here, we follow a similar analysis to the previous description for the rotational bands of Carbon-12 and its Hoyle state \cite{Lau:2014baa}. For Oxygen-16, the ground state is $0^+$ and the first $3^-$ state is lower in energy than the first $2^+$ state. This is the signature of the rotational spectrum of a tetrahedral object. In the Skyrme model, several $B=16$ solutions are known. They are constructed from four $B=4$ cubic Skyrmions arranged in a bent square ($D_{2d}$), flat square ($D_{4h}$) and tetrahedral ($T_d$) configuration, respectively (see Fig.~\ref{Sky_bary_dens} (e)(f)(g)). Hence, we interpret the ground state band in terms of the tetrahedral Skyrmion. The quantized $T_d$-symmetric Skyrmion models the ground state of Oxygen-16 and its $3^-$ and $4^+$ rotational excitations.

The bent and flat square Skyrmions have similar energy and can be seen as energy degenerate within the limits of our numerical accuracy.  Note that we cannot confirm the result of the article \cite{Battye:2006na} that the flat square is of noticeable higher energy than the bent square. 

For the $B=16$ $D_{4h}$-symmetric, flat square Skyrmion, the Finkelstein--Rubinstein (F-R) constraints on a wavefunction $\psi$ with zero isospin are
\begin{eqnarray}\label{flat16_FR}
e^{\, i \frac{\pi}{2} \hat{L}_3} \left|\psi\right\rangle = \left|\psi\right\rangle\,\quad \text{and }\quad e^{\, i \pi \hat{L}_1} \left|\psi\right\rangle = \left|\psi\right\rangle\,,
\end{eqnarray}
where $\hat{L}_i$ is the spin operator projected on the body-fixed $i$th axis. $k$ is the eigenvalue of $\hat{L}_3$. There is a $k=0$ rotational band, but the first constraint excludes states with $k = 2$. For $J=2$,  the only state allowed by the $D_{4h}$ symmetry is $\left|2\,, 0\right\rangle$. The parity operator of this Skyrmion quantized with zero isospin is the identity operator. Hence, all  the states have positive parity. This misses important states in the Oxygen-16 spectrum, so we turn to the bent square.

For the $B=16$ $D_{2d}$-symmetric, bent square Skyrmion, the F-R constraints are
\begin{eqnarray}\label{bent16_FR}
e^{\,i \pi \hat{L}_3} \left|\psi\right\rangle =\left|\psi\right\rangle\,\quad \text{and }\quad e^{\,i \pi\hat{L}_1} \left|\psi\right\rangle = \left|\psi\right\rangle\,,
\end{eqnarray}
and the parity operator is
\begin{equation}
\hat{P}=e^{i \frac{\pi}{2} \hat{L}_3} \,.
\end{equation}
In this case, there is a rotational band with $k=0$ and a band with $k=2$, and there are two $J=2$ states, $\left|2\,, 0\right\rangle$ and $\frac{1}{\sqrt{2}}\left(\left|2\,, 2\right\rangle + \left|2\,,-\!2\right\rangle\right)$, with $k=0$ and $k=2$ respectively. The states in the $k=0$ band have positive parity while the states in the $k=2$ band have negative parity. These bands can be identified with $0^+, 2^+, 4^+$ and $2^-,3^-,4^-$ states in the experimentally measured spectrum. Hence, the quantized $B=16$ bent square is preferable for modelling the second excited spin-0 state $0_2^{+}$ of Oxygen-16 and its rotational excitations.

\begin{figure}[!h]
\centering
\includegraphics[totalheight=9.cm]{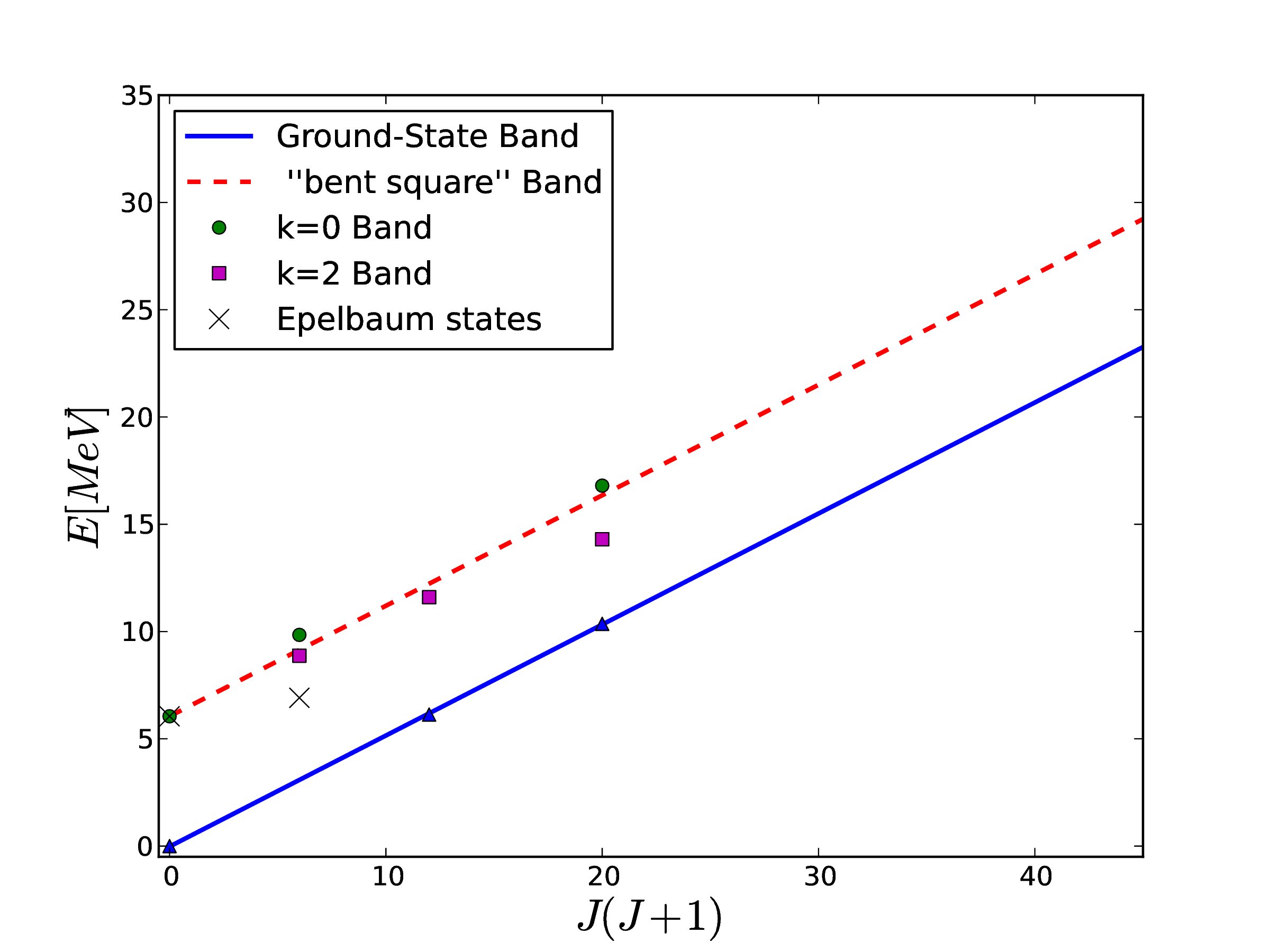}
\caption{Experimental states of Oxygen-16. The symbol triangle denotes the states of the ground state band, and circle and square denote the $k=0,2$ states of the ``bent square'' bands. The $0_1^+, 3^-, 4^+$ states in the ground state band have energies $0.0$, $6.13$ and $10.36$ MeV. For the ``bent square'' band, the $0_2^+,2^+,4^+$ states with $k=0$ have energies $6.05$, $9.84$ and $16.84$ MeV and the $2^-,3^-,4^-$ states with $k=2$ have energies $8.87$, $11.60$ and $14.30$ MeV. The symbol cross represents the $2_1^+$ state which has been interpreted as a rotational excitation of the $0^+_2$ state by Epelbaum \textit{et. al.} ~\cite{Epelbaum:2013paa}. See the text for more details.}
\label{B16_rotband}
\end{figure}

To test further this identification we can use similar techniques as applied in Ref.~\cite{Lau:2014baa} to the rotational excitations of Carbon-12 and its Hoyle state. All $B=16$ Skyrmion solutions have inertia tensors of symmetric-top type with $V_{11}=V_{22}$ and $V_{33}$ the same or distinct. For the tetrahedral solution we find $V_{11}=V_{22}=9100$ and $V_{33}=9128$, where the difference must be a numerical artifact. For the bent square, $V_{11}=V_{22}=9123$ and $V_{33}=14602$.

The energy eigenvalues of the quantum Hamiltonian for purely rotational motion of a symmetric top are given by
\begin{eqnarray}
E(J,k)=C\left\{\frac{1}{2V_{11}}J(J+1)+\left(\frac{1}{2V_{33}}-\frac{1}{2V_{11}}\right)k^2\right\}\,,
\label{En_JK}
\end{eqnarray}
where $J$ denotes the total spin quantum number, $k$ is the eigenvalue of $\hat{L}_3$, and $C$ is a dimensional conversion factor from Skyrme to physical units \cite{Lau:2014baa}. Here, $C$ is a purely phenomenological parameter as has previously been used in the discussion of Carbon-12 and its rotational states in Ref.~\cite{Lau:2014baa}. In Fig.~\ref{B16_rotband}, we plot against $J(J+1)$ the energies of experimentally observed Oxygen-16 states up to spin 4 in the ground-state band and in the rotational bands formed by the rotational excitations of the $0_2^+$ state and the $2^-$ state. Taking $C$ as our fitting parameter, we fit Eq.~(\ref{En_JK}) to the $0^+,\,3^-,\,4^+$ states of the ground-state band, whose energies are $0.0$, $6.13$ and $10.36$ MeV. The linear fit gives $C=9418$ MeV, and therefore a best fit slope of $0.517$ MeV, using the $V_{11}$ value $9100$ for the tetrahedral Skyrmion. 

We find that the experimental slope of the $k=0$ band based on the $0_2^+$ state then agrees very well with the Skyrme model prediction. Eq.~(\ref{En_JK}) gives a theoretical slope of $0.516$ MeV, where we used the bent square's moment of inertia $V_{11}=9123$ and $C=9418$ MeV as derived above. The experimental slope is estimated from the best linear fit to the $0_2^+,2^+,4^+$ states with energies $6.05,9.84$ and 16.84 MeV to be $0.545$ MeV. The Skyrme model prediction for the ratio of the slopes is just the ratio of the $V_{11}$ values for the $D_{2d}$- and $T_d$-symmetric Skyrmions, which is $9123/9100=1.00$. Note that the dimensional conversion factor $C$ cancels. For comparison, the experimental ratio of the slopes is $1.05$. 

We also include in Fig.~\ref{B16_rotband} the experimental $2^-,3^-,4^-$ states of energies $8.87,11.60$ and 14.30 MeV which we interpret as the $k=2$ band formed by the rotational excitations of the bent square. The $k=2$ band lies below the $k=0$ band, agreeing with the oblateness of the bent square Skyrmion. For an oblate configuration, $V_{11} < V_{33}$, and according to Eq. (\ref{En_JK}), for a fixed spin $J$, the energy of a state with non-zero $k$ has lower energy than a $k=0$ state. The predicted energy difference between states with the same spin $J$ is $0.77$ MeV between the $k=0$ and $k=2$ bands. This agrees marginally with the experimentally measured differences of $0.97$ MeV for the spin-2 states and $2.54$ MeV for the spin-4 states.

We calculate the $E2$ transition strength from the $0_2^{+}$ state of energy $6.05$ MeV to the $2^{+}$ state of energy 9.84 MeV by taking the bent square Skyrmion as the underlying structure. We obtain $B(E2:0_2^{+}\rightarrow2^{+})=0.0168 \,\text{e}^2\text{b}^2$. The associated intrinsic, oblate quadrupole moment is $Q_0=-0.412\,\text{e}\text{b}$. For this transition, we are unable to find experimental $E2$ values in the literature. For completeness, we also include in Tables~\ref{Tab_Q} and \ref{Tab_BE2} the quadrupole moment and $B(E2)$ values when modelling the $0_2^{+}$ state and its spin-2 excitation $2^{+}$ by the flat square Skyrmion.

Epelbaum et al. \cite{Epelbaum:2013paa} have considered the transition between the $0_2^{+}$ state and the lowest spin-2 state $2_1^{+}$ of energy $6.91$ MeV. These states are represented in Fig.~\ref{B16_rotband} by the cross symbols. They interpret the  $2_1^{+}$ state as a rotational excitation of a square configuration of alpha clusters. However, it is badly described as a rotational excitation of the bent or flat square $B=16$ Skyrmion. The $B(E2:2_1^{+}\rightarrow0_2^{+})$ transition strength predicted in Ref.~\cite{Epelbaum:2013paa} is based on nuclear lattice effective field theory simulations. The up transition strength is $B(E2:0_2^{+}\rightarrow2_1^{+})=0.0110\,\text{e}^2\text{b}^2$. The corresponding empirical value is found to be $B(E2:0_2^{+}\rightarrow2_1^{+})=0.0325\,\text{e}^2\text{b}^2$ \cite{AjzenbergSelove19711,Epelbaum:2013paa}.

\section{Nonzero Isospin}

In the previous sections, we restricted the discussion to $E2$ transitions in the absence of isospin. In this section, we show how quadrupole transition strengths can be calculated within the Skyrme model in the presence of nonzero nuclear isospin.

\subsubsection{Beryllium-12}

Beryllium-12 is a nucleus in an $I=2$ isospin multiplet, as it has four protons and eight neutrons. Nuclei of mass number 12, with isospin 0, 1, and 2  are especially well described within the Skyrme model as quantum states of the $D_{3h}$-symmetric $B=12$ Skyrmion \cite{Battye:2009ad} (see baryon density isosurface displayed in Fig.~\ref{Sky_bary_dens} (d)). In particular, the low-lying energy levels of Beryllium-12, with various spins, appear as states with $I=2$ and $I_3 = -2$.  

The quantum states $\left|\psi_{J^\pi,I,|L_3|,|K_3|}\rangle\right.$ of the $B = 12$ Skyrmion allowed by the Finkelstein-Rubinstein constraints \cite{Krusch:2002by,Krusch:2005iq} are listed in Table~6 of Ref.~\cite{Battye:2009ad}. $J$ and $I$ are the total spin and isospin labels, and $|L_3|$ and $|K_3|$ are the projections on to ``body-fixed'' 3-axes. Both for $J=0$ and $J=2$ there is a unique allowed $I=2$ state, with $K_3 = 0$. The ``space-fixed'' isospin $I_3$ can take any integer value from $-2$ to $2$, and for Beryllium-12 it is $I_3 =-2$. Suppressing the spin state, we denote the isospin state of Beryllium-12 as $|2,0;-2\rangle$.

Beryllium-12 exhibits a large quadrupole strength in the transition between the $0^+$ ground state and the first $2^+$ state at $2.1$ MeV \cite{Imai:2009zza}. In Ref.~\cite{Imai:2009zza}, a $B(E2:2^+\rightarrow 0^+)$ value of $0.0008\,\text{e}^2\text{b}^2$ has been determined through the lifetime measurement of the $2^+$ state. Using Eq.~(\ref{Trans_up}) this results in a $B(E2:0^+\rightarrow 2^+)$ value of $0.0040\,\text{e}^2\text{b}^2$.

Here, we consider $E2$ transitions between these spin states in the Skyrme model, using the Skyrmion's isospin state $|2,0;-2\rangle$. The new aspect is to take account of the contribution of the isospin to the electric charge density, and hence to the quadrupole moments.

It would be best to do a proper quantum calculation of the expectation value of the quadrupole moments, but we have not been able to do this. Instead we treat the isospin state using a classical approximation. This is analogous to the approach to nucleons adopted in \cite{Manton:2011mi,Foster:2015cpa}. There a nucleon is treated as a classically spinning $B=1$ Skyrmion, which gives it both spin and isospin. To achieve the desired  ``space-fixed'' isospin projection, the $B=12$ Skyrmion's red/green/blue colours spin in isospace while the black/white colours do not. (This produces a rotation among the $\pi_1$ and $\pi_2$ fields, while the $\pi_3$ field, associated with the black/white axis, remains constant.) The role of the isospin state $|2,0;-2\rangle$ is to inform us of the most likely colouring of the Skyrmion, before the colours spin. If the projection in this state was $K_3 = \pm 2$ then the standard orientation and colouring of the Skyrmion would be the correct one, but as the projection is $K_3 = 0$ we must reorientate the colourings first. 

We reorientate the classical $B=12$ Skyrmion colouring by the angles that maximize the wavefunction $|2,0;-2\rangle$.  The associated Wigner $\mathcal{D}$ function for this state takes the form $\mathcal{D}^2_{0,-2}(\alpha,\beta,\gamma)=e^{-2i\gamma}\sin^2\beta$, where $\alpha, \beta, \gamma$ are the isorotational Euler angles. The factor $e^{-2i\gamma}$ is the quantum representation of the colours spinning with $I_3 = -2$. $e^{-2i\gamma}\sin^2\beta$ has its maximum magnitude at $\beta=\pi/2$ and $\gamma=0$ (or any other value of $\gamma$). Hence, we perform an isospin rotation of our initial classical solution with $\beta=\pi/2$ and $\gamma=0$. This rotates the black/white points of the Skyrmion to be on the faces or edges of the $B=4$ cube constituents, rather than at the vertices. 

In detail, under an isospin rotation with $\beta=\pi/2$ and $\gamma=0$ (and $\alpha$ undetermined) the pion field $\boldsymbol{\pi}=(\pi_1,\pi_2,\pi_3)$ transforms to
\begin{eqnarray}
\pi_1^\prime=-\sin\alpha\ \pi_2+\cos\alpha\ \pi_3\,,\quad\pi_2^\prime=\cos\alpha\ \pi_2+\sin\alpha\ \pi_3\,,\quad \pi_3^\prime=-\pi_1\,,
\end{eqnarray}
and hence the new moment of inertia that we need is $U_{33}^\prime=U_{11}$. The colours spin about the new 3-axis in isospace, but dynamically this is equivalent to spinning about the old 1-axis. For the $D_{3h}$-symmetric $B=12$ Skyrmion we find $U_{33}^\prime = U_{11}=442$, see Table~\ref{Tab_Sky_inertia}. To classically model a Beryllium-12 nucleus whose projected isospin has magnitude $-2$ we require $U_{33}^\prime\omega=-2\hbar$. (Physical isospin, like spin, is a half-integer or integer multiple of $\hbar$.)

Hence, the isorotational angular velocity $\omega$ is given by 
\begin{eqnarray}\label{Be12_omega}
\omega=-\frac{2\hbar}{U_{33}^\prime}=-0.14\,,
\end{eqnarray}
where $\hbar=30.2$. For the Beryllium-12 nucleus, the classical 
isospin density is
\begin{eqnarray}\label{Be12_IsoDens}
\mathcal{I}_3(\mathbf{x})=\frac{\omega}{\hbar}\mathcal{U}_{33}^\prime(\mathbf{x})\,.
\end{eqnarray}
This isospin contribution to the electric charge density (\ref{Sky_current}) has to be taken into account when calculating quadrupole moments (\ref{Sky_quadrupole}). Note that Eq.~(\ref{Be12_IsoDens}) is correctly normalized as its integral gives $I_3=-2$ and decreases the total electric charge $\frac{1}{2}B+I_3$ of the $B=12$ Skyrmion from $6$ to $4$.

We compute numerically the electric quadrupole moments $Q_{11}^{\text{Sky}}=  5.65$, $Q_{22}^{\text{Sky}}=6.32$ and $Q_{33}^{\text{Sky}}=-11.9$ for the reorientated $B=12$ Skyrmion. Thus, expressed in physical units using the calibration (\ref{Cali_LM}) the intrinsic quadrupole is $Q_{0}=-0.135\,\text{e}\text{b}$. The corresponding $B(E2:0\rightarrow 2)$ value is $0.00181\, \text{e}^2\text{b}^2$ which differs by approximately $50\%$ from the experimental value.

\section{Conclusions}

We have calculated for the first time within the Skyrme model electromagnetic transition strengths between the $0^+$ ground state and the first-excited $2^+$ state for a range of light nuclei: \ce{^{8}_{4} Be_4}, \ce{^{12}_{6} C_6} and its Hoyle state, \ce{^{12}_{4} Be_8}, \ce{^{20}_{10} Ne_{10}} and \ce{^{24}_{12} Mg_{12}}. We find that the calculated $E2$ transition strengths have the correct order of magnitude and the computed intrinsic quadrupole moments match the experimentally observed effective nuclear shapes. For the Hoyle state we predict a large $B(E2)\!\uparrow$ value of  $0.0521\, \text{e}^2\text{b}^2$. Measurements of the electromagnetic transition strengths between the states of the Hoyle band are technically difficult and have yet to be performed \cite{Freer:2014qoa,2014LNP...875...25J}. 

For Oxygen-16, we can obtain a quantitative understanding of the ground state band and the rotational band formed by the second excited spin-0 state $0_2^+$ and its rotational excitations. Similarly to the ground state band of Carbon-12 and the rotational band of the Hoyle state \cite{Lau:2014baa}, we interpret the Oxygen-16 rotational bands as rotational excitations of two Skyrmions with very different shapes, one tetrahedral and the other a bent square. The quantized tetrahedral Skyrmion models the $0_1^+$ ground state and  its $3^-$ and $4^+$ excitations. The quantized bent square Skyrme configuration is identified with the $0_2^+$ state and its rotational excitations. We find that the $0_2^+,2^+,4^+$ states of energies $6.05,9.84$ and 16.84 MeV are very well modelled as $k=0$ states of the bent square band. The almost equal values of the spin moment of inertia $V_{11}$ for the tetrahedron and bent square are a success of the Skyrme model. The Skyrme model predicts that the ratio of the slopes of the $k=0$ bent square band and the ground state band is the ratio of these $V_{11}$ values, and is very 
close to 1. The ratio of the experimental slopes agrees with this. The $k=2$ band of the bent square matches experimental $2^-,3^-,4^-$ states. Furthermore, we used Eq. (\ref{En_JK}) to predict the energy difference between $k=0$ and $k=2$ states of the same spin (but opposite parity). The predicted energy difference has the right sign and marginally agrees with experiment. 

There remain some challenges for the Skyrme model. For baryon number $20$, the triangular bipyramidal arrangement of five $B=4$ cubes which we used to describe $E2$ transitions in Neon-20 is not a minimal energy Skyrmion but a saddle point. For $B=32$, the minimal energy Skyrmion is cubically symmetric and hence cannot explain the large prolate quadrupole moment of Sulphur-32. However, Skyrmions deform under rotations and hence a non-zero quadrupole moment may be induced.

The approach used in this paper is limited to Skyrmions with axially symmetric inertia tensors, and with quadrupole moments satisfying $Q_{11} = Q_{22}$. The calculation of $B(E2)$ strengths for transitions between rotational levels in triaxial nuclei \cite{0022-3689-2-4-006,PhysRevC.78.014302} using the Skyrme model requires a different approach. Further lines of investigation to consider are higher-order electric multipole transitions and magnetic dipole transitions within the Skyrme model. In particular, the $0^+$ to $3^-$ transition strength in the Oxygen-16 ground state band should be calculated. Finally, we neglected deformations in our calculations; that is, we assumed that the low-lying rotational states are well approximated by the rigid rotor states of the Skyrmions. Recently, $E2$ transitions in deformed nuclei have been studied within an effective theory for axially symmetric systems  \cite{Perez:2015tta}, and a similar study of deformed Skyrmions is desirable.

\section*{Acknowledgements}

This work was partly undertaken on the COSMOS Shared Memory system at DAMTP, University of Cambridge operated on behalf of the STFC DiRAC HPC Facility. This equipment is funded by BIS National E-infrastructure capital grant ST/J005673/1 and STFC grants ST/J001341/1, ST/H008586/1, ST/K00333X/1. MH has been partially funded by the UK Science and Technology Facilities Council under grant number ST/J000434/1. MH thanks Andrzej Wereszczynski and the Jagiellonian University, Krakow for hospitality. PHCL thanks Ling-Yan Hung and Fudan University in Shanghai for hospitality. PHCL acknowledges support as an International Research Fellow of the Japan Society for the Promotion of Science (JSPS).


\begin{thebibliography}{58}
\expandafter\ifx\csname natexlab\endcsname\relax\def\natexlab#1{#1}\fi
\expandafter\ifx\csname bibnamefont\endcsname\relax
  \def\bibnamefont#1{#1}\fi
\expandafter\ifx\csname bibfnamefont\endcsname\relax
  \def\bibfnamefont#1{#1}\fi
\expandafter\ifx\csname citenamefont\endcsname\relax
  \def\citenamefont#1{#1}\fi
\expandafter\ifx\csname url\endcsname\relax
  \def\url#1{\texttt{#1}}\fi
\expandafter\ifx\csname urlprefix\endcsname\relax\def\urlprefix{URL }\fi
\providecommand{\bibinfo}[2]{#2}
\providecommand{\eprint}[2][]{\url{#2}}

\bibitem[{\citenamefont{Bohr and Mottelson}(1998{\natexlab{a}})}]{bohrMot1}
\bibinfo{author}{\bibfnamefont{A.}~\bibnamefont{Bohr}} \bibnamefont{and}
  \bibinfo{author}{\bibfnamefont{B.~R.} \bibnamefont{Mottelson}},
  \emph{\bibinfo{title}{Nuclear Structure, Volume I: Single Particle Motion}}
  (\bibinfo{publisher}{World Scientific}, \bibinfo{address}{Singapore},
  \bibinfo{year}{1998}{\natexlab{a}}).



\bibitem[{\citenamefont{Bohr and Mottelson}(1998{\natexlab{b}})}]{bohr1998nuclear2}
\bibinfo{author}{\bibfnamefont{A.}~\bibnamefont{Bohr}} \bibnamefont{and}
  \bibinfo{author}{\bibfnamefont{B.~R.} \bibnamefont{Mottelson}},
  \emph{\bibinfo{title}{Nuclear Structure, Volume II: Nuclear Deformations}}
  (\bibinfo{publisher}{World Scientific}, \bibinfo{address}{Singapore},
  \bibinfo{year}{1998}{\natexlab{b}}).



\bibitem[{\citenamefont{{Jenkins}}(2014)}]{2014LNP...875...25J}
\bibinfo{author}{\bibfnamefont{D.~G.} \bibnamefont{{Jenkins}}}, in
  \emph{\bibinfo{booktitle}{Lecture Notes in Physics}}, edited by

  \bibinfo{editor}{\bibfnamefont{C.}~\bibnamefont{{Beck}}}
  (\bibinfo{publisher}{Springer Verlag}, \bibinfo{address}{Berlin},
  \bibinfo{year}{2014}), vol. \bibinfo{volume}{875} of
  \emph{\bibinfo{series}{Lecture Notes in Physics}}, p.~\bibinfo{pages}{25}.



\bibitem[{\citenamefont{Bhatt et~al.}(1992)\citenamefont{Bhatt, Nestor, and Raman}}]{PhysRevC.46.164}
\bibinfo{author}{\bibfnamefont{K.~H.} \bibnamefont{Bhatt}},
  \bibinfo{author}{\bibfnamefont{C.~W.} \bibnamefont{Nestor}},
  \bibnamefont{and} \bibinfo{author}{\bibfnamefont{S.}~\bibnamefont{Raman}},
  \bibinfo{journal}{Phys. Rev.} \textbf{\bibinfo{volume}{C46}},
  \bibinfo{pages}{164} (\bibinfo{year}{1992}).



\bibitem[{\citenamefont{Raman et~al.}(2001)\citenamefont{Raman, Nestor, and Tikkanen}}]{Raman:1201zz}
\bibinfo{author}{\bibfnamefont{S.}~\bibnamefont{Raman}},
  \bibinfo{author}{\bibfnamefont{C.~W.} \bibnamefont{Nestor}},
  \bibnamefont{and} \bibinfo{author}{\bibfnamefont{P.}~\bibnamefont{Tikkanen}},
  \bibinfo{journal}{At. Data Nucl. Data Tables} \textbf{\bibinfo{volume}{78}},
  \bibinfo{pages}{1} (\bibinfo{year}{2001}).



\bibitem[{\citenamefont{Pritychenko et~al.}(2014)\citenamefont{Pritychenko, Birch, Horoi, and Singh}}]{Pritychenko:2013taa}
\bibinfo{author}{\bibfnamefont{B.}~\bibnamefont{Pritychenko}},
  \bibinfo{author}{\bibfnamefont{M.}~\bibnamefont{Birch}},
  \bibinfo{author}{\bibfnamefont{M.}~\bibnamefont{Horoi}}, \bibnamefont{and}
  \bibinfo{author}{\bibfnamefont{B.}~\bibnamefont{Singh}},
  \bibinfo{journal}{Nucl. Data Sheets} \textbf{\bibinfo{volume}{120}},
  \bibinfo{pages}{111} (\bibinfo{year}{2014}).



\bibitem[{\citenamefont{Skyrme}(1961)}]{Skyrme:1961vq}
\bibinfo{author}{\bibfnamefont{T.~H.~R.} \bibnamefont{Skyrme}},
  \bibinfo{journal}{Proc. Roy. Soc. Lond.} \textbf{\bibinfo{volume}{A260}},
  \bibinfo{pages}{127} (\bibinfo{year}{1961}).



\bibitem[{\citenamefont{Skyrme}(1962)}]{Skyrme:1962vh}
\bibinfo{author}{\bibfnamefont{T.~H.~R.} \bibnamefont{Skyrme}},
  \bibinfo{journal}{Nucl. Phys.} \textbf{\bibinfo{volume}{31}},
  \bibinfo{pages}{556} (\bibinfo{year}{1962}).


\bibitem[{\citenamefont{Blatt and Weisskopf}(1952)}]{blattWeiss}
\bibinfo{author}{\bibfnamefont{J.}~\bibnamefont{Blatt}} \bibnamefont{and}
  \bibinfo{author}{\bibfnamefont{V.}~\bibnamefont{Weisskopf}},
  \emph{\bibinfo{title}{{Theoretical Nuclear Physics}}}
  (\bibinfo{publisher}{Wiley}, \bibinfo{address}{New York},
  \bibinfo{year}{1952}).



\bibitem[{\citenamefont{Brink et~al.}(1970)\citenamefont{Brink, Friedrich, Weiguny, and Wong}}]{Brink1970143}
\bibinfo{author}{\bibfnamefont{D.}~\bibnamefont{Brink}},
  \bibinfo{author}{\bibfnamefont{H.}~\bibnamefont{Friedrich}},
  \bibinfo{author}{\bibfnamefont{A.}~\bibnamefont{Weiguny}}, \bibnamefont{and}
  \bibinfo{author}{\bibfnamefont{C.}~\bibnamefont{Wong}},
  \bibinfo{journal}{Phys. Lett.} \textbf{\bibinfo{volume}{B33}},
  \bibinfo{pages}{143} (\bibinfo{year}{1970}).



\bibitem[{\citenamefont{Battye et~al.}(2007)\citenamefont{Battye, Manton, and  Sutcliffe}}]{Battye:2006na}
\bibinfo{author}{\bibfnamefont{R.~A.} \bibnamefont{Battye}},
  \bibinfo{author}{\bibfnamefont{N.~S.} \bibnamefont{Manton}},
  \bibnamefont{and} \bibinfo{author}{\bibfnamefont{P.~M.}
  \bibnamefont{Sutcliffe}}, \bibinfo{journal}{Proc. Roy. Soc. Lond.}
  \textbf{\bibinfo{volume}{A463}}, \bibinfo{pages}{261} (\bibinfo{year}{2007}).



\bibitem[{\citenamefont{Feist et~al.}(2013)\citenamefont{Feist, Lau, and Manton}}]{Feist:2012ps}
\bibinfo{author}{\bibfnamefont{D.~T.~J.} \bibnamefont{Feist}},
  \bibinfo{author}{\bibfnamefont{P.~H.~C.} \bibnamefont{Lau}},
  \bibnamefont{and} \bibinfo{author}{\bibfnamefont{N.~S.}
  \bibnamefont{Manton}}, \bibinfo{journal}{Phys. Rev.}
  \textbf{\bibinfo{volume}{D87}}, \bibinfo{pages}{085034}
  (\bibinfo{year}{2013}).



\bibitem[{\citenamefont{Krusch}(2003)}]{Krusch:2002by}
\bibinfo{author}{\bibfnamefont{S.}~\bibnamefont{Krusch}},
  \bibinfo{journal}{Ann. Phys.} \textbf{\bibinfo{volume}{304}},
  \bibinfo{pages}{103} (\bibinfo{year}{2003}).



\bibitem[{\citenamefont{Krusch}(2006)}]{Krusch:2005iq}
\bibinfo{author}{\bibfnamefont{S.}~\bibnamefont{Krusch}},
  \bibinfo{journal}{Proc. Roy. Soc. Lond.} \textbf{\bibinfo{volume}{A462}},
  \bibinfo{pages}{2001} (\bibinfo{year}{2006}).



\bibitem[{\citenamefont{Manko et~al.}(2007)\citenamefont{Manko, Manton, and Wood}}]{Manko:2007pr}
\bibinfo{author}{\bibfnamefont{O.~V.} \bibnamefont{Manko}},
  \bibinfo{author}{\bibfnamefont{N.~S.} \bibnamefont{Manton}},
  \bibnamefont{and} \bibinfo{author}{\bibfnamefont{S.~W.}~\bibnamefont{Wood}},
  \bibinfo{journal}{Phys. Rev.} \textbf{\bibinfo{volume}{C76}},
  \bibinfo{pages}{055203} (\bibinfo{year}{2007}).



\bibitem[{\citenamefont{Battye et~al.}(2009)\citenamefont{Battye, Manton, Sutcliffe, and Wood}}]{Battye:2009ad}
\bibinfo{author}{\bibfnamefont{R.~A.} \bibnamefont{Battye}},
  \bibinfo{author}{\bibfnamefont{N.~S.} \bibnamefont{Manton}},
  \bibinfo{author}{\bibfnamefont{P.~M.} \bibnamefont{Sutcliffe}},
  \bibnamefont{and} \bibinfo{author}{\bibfnamefont{S.~W.} \bibnamefont{Wood}},
  \bibinfo{journal}{Phys. Rev.} \textbf{\bibinfo{volume}{C80}},
  \bibinfo{pages}{034323} (\bibinfo{year}{2009}).



\bibitem[{\citenamefont{Lau and Manton}(2014{\natexlab{a}})}]{Lau:2014sva}
\bibinfo{author}{\bibfnamefont{P.~H.~C.} \bibnamefont{Lau}} \bibnamefont{and}
  \bibinfo{author}{\bibfnamefont{N.~S.} \bibnamefont{Manton}},
  \bibinfo{journal}{Phys. Rev.} \textbf{\bibinfo{volume}{D89}},
  \bibinfo{pages}{125012} (\bibinfo{year}{2014}{\natexlab{a}}).



\bibitem[{\citenamefont{Lau and Manton}(2014{\natexlab{b}})}]{Lau:2014baa}
\bibinfo{author}{\bibfnamefont{P.~H.~C.} \bibnamefont{Lau}} \bibnamefont{and}
  \bibinfo{author}{\bibfnamefont{N.~S.} \bibnamefont{Manton}},
  \bibinfo{journal}{Phys. Rev. Lett.} \textbf{\bibinfo{volume}{113}},
  \bibinfo{pages}{232503} (\bibinfo{year}{2014}{\natexlab{b}}).

\bibitem[{\citenamefont{Adkins et~al.}(1983)\citenamefont{Adkins, Nappi, and Witten}}]{Adkins:1983ya}
\bibinfo{author}{\bibfnamefont{G.}~\bibnamefont{Adkins}},
  \bibinfo{author}{\bibfnamefont{C.}~\bibnamefont{Nappi}}, \bibnamefont{and}
  \bibinfo{author}{\bibfnamefont{E.}~\bibnamefont{Witten}},
  \bibinfo{journal}{Nucl. Phys.} \textbf{\bibinfo{volume}{B228}},
  \bibinfo{pages}{552} (\bibinfo{year}{1983}).

\bibitem[{\citenamefont{Braaten et~al.}(1988)\citenamefont{Braaten and Carson}}]{Braaten:1988cc}
\bibinfo{author}{\bibfnamefont{E.}~\bibnamefont{Braaten}} \bibnamefont{and}
  \bibinfo{author}{\bibfnamefont{L.}~\bibnamefont{Carson}},
  \bibinfo{journal}{Phys. Rev.} \textbf{\bibinfo{volume}{D38}},
  \bibinfo{pages}{3525} (\bibinfo{year}{1988}).
  
\bibitem[{\citenamefont{Kopeliovich}(2001)}]{Kopeliovich:2001yg}
\bibinfo{author}{\bibfnamefont{V.~B.} \bibnamefont{Kopeliovich}},
  \bibinfo{journal}{JETP} \textbf{\bibinfo{volume}{93}},
  \bibinfo{pages}{435} (\bibinfo{year}{2001}).


\bibitem[{\citenamefont{Manton and Sutcliffe}(2004)}]{Manton:2004tk}
\bibinfo{author}{\bibfnamefont{N.~S.} \bibnamefont{Manton}} \bibnamefont{and}
  \bibinfo{author}{\bibfnamefont{P.~M.} \bibnamefont{Sutcliffe}},
  \emph{\bibinfo{title}{{Topological Solitons}}} (\bibinfo{publisher}{Cambridge
  Univ. Press}, \bibinfo{address}{Cambridge}, \bibinfo{year}{2004}).



\bibitem[{\citenamefont{Brown and Rho~(eds.)}(2010)}]{Brown:2009eh}
\bibinfo{author}{\bibfnamefont{G.~E.} \bibnamefont{Brown}} \bibnamefont{and}
  \bibinfo{author}{\bibfnamefont{M.}~\bibnamefont{Rho~(eds.)}},
  \emph{\bibinfo{title}{{The Multifaceted Skyrmion}}}
  (\bibinfo{publisher}{World Scientific}, \bibinfo{address}{Singapore},
  \bibinfo{year}{2010}).



\bibitem[{\citenamefont{Gisiger and Paranjape}(1998)}]{Gisiger:1998tv}
\bibinfo{author}{\bibfnamefont{T.}~\bibnamefont{Gisiger}} \bibnamefont{and}
  \bibinfo{author}{\bibfnamefont{M.~B.} \bibnamefont{Paranjape}},
  \bibinfo{journal}{Phys. Rept.} \textbf{\bibinfo{volume}{306}},
  \bibinfo{pages}{109} (\bibinfo{year}{1998}).


\bibitem[{\citenamefont{Adkins and Nappi}(1984)}]{Adkins:1983hy}
\bibinfo{author}{\bibfnamefont{G.}~\bibnamefont{Adkins}} \bibnamefont{and}
  \bibinfo{author}{\bibfnamefont{C.}~\bibnamefont{Nappi}},
  \bibinfo{journal}{Nucl. Phys.} \textbf{\bibinfo{volume}{B233}},
  \bibinfo{pages}{109} (\bibinfo{year}{1984}).


\bibitem[{\citenamefont{Shewchuk}(1994)}]{NCG}
\bibinfo{author}{\bibfnamefont{J.~R.} \bibnamefont{Shewchuk}},
  \emph{\bibinfo{title}{An introduction to the conjugate gradient method without the agonizing pain}} (\bibinfo{year}{1994}),
  \urlprefix\url{http://www.cs.cmu.edu/~quake-papers/painless-conjugate-gradient.pdf}.



\bibitem[{\citenamefont{Feist}(2012)}]{Feist:2011aa}
\bibinfo{author}{\bibfnamefont{D.~T.~J.} \bibnamefont{Feist}},
  \bibinfo{journal}{JHEP} \textbf{\bibinfo{volume}{1202}}, \bibinfo{pages}{100}
  (\bibinfo{year}{2012}).



\bibitem[{\citenamefont{Battye and Sutcliffe}(2002)}]{Battye:2001qn}
\bibinfo{author}{\bibfnamefont{R.~A.} \bibnamefont{Battye}} \bibnamefont{and}
  \bibinfo{author}{\bibfnamefont{P.~M.} \bibnamefont{Sutcliffe}},
  \bibinfo{journal}{Rev. Math. Phys.} \textbf{\bibinfo{volume}{14}},
  \bibinfo{pages}{29} (\bibinfo{year}{2002}).


  \bibitem[{\citenamefont{Kopeliovich}(1988)\citenamefont{Kopeliovich}}]{Kopeliovich:1988np}
\bibinfo{author}{\bibfnamefont{V.~B.}~\bibnamefont{Kopeliovich}},
  \bibinfo{journal}{Sov. J. Nucl. Phys.} \textbf{\bibinfo{volume}{47}},
  \bibinfo{pages}{949} (\bibinfo{year}{1988}).  
  
  
\bibitem[{\citenamefont{Battye et~al.}(2014)\citenamefont{Battye, Haberichter, and Krusch}}]{Battye:2014qva}
\bibinfo{author}{\bibfnamefont{R.~A.} \bibnamefont{Battye}},
  \bibinfo{author}{\bibfnamefont{M.}~\bibnamefont{Haberichter}},
  \bibnamefont{and} \bibinfo{author}{\bibfnamefont{S.}~\bibnamefont{Krusch}},
  \bibinfo{journal}{Phys. Rev.} \textbf{\bibinfo{volume}{D90}},
  \bibinfo{pages}{125035} (\bibinfo{year}{2014}).

\bibitem[{\citenamefont{Manton}(2012)}]{Manton:2011mi}
\bibinfo{author}{\bibfnamefont{N.~S.} \bibnamefont{Manton}},
  \bibinfo{journal}{Math. Method. Appl. Sci.} \textbf{\bibinfo{volume}{35}},
  \bibinfo{pages}{1188} (\bibinfo{year}{2012}).



\bibitem[{\citenamefont{Tanihata et~al.}(1985)}]{Tanihata:1986kh}
\bibinfo{author}{\bibfnamefont{I.}~\bibnamefont{Tanihata}}
  \bibnamefont{et~al.}, \bibinfo{journal}{Phys. Rev. Lett.}
  \textbf{\bibinfo{volume}{55}}, \bibinfo{pages}{2676} (\bibinfo{year}{1985}).




\bibitem[{\citenamefont{Schwesinger and Weigel}(2015)}]{Schwesinger:1991fy}
\bibinfo{author}{\bibfnamefont{B.} \bibnamefont{Schwesinger}}
  \bibnamefont{and}
  \bibinfo{author}{\bibfnamefont{H.}~\bibnamefont{Weigel}},
  \bibinfo{journal}{Phys. Lett.} \textbf{\bibinfo{volume}{B267}},
  \bibinfo{pages}{438} (\bibinfo{year}{1991}).



\bibitem[{\citenamefont{Weigel}(1996)}]{Weigel:1995cz}
  \bibinfo{author}{\bibfnamefont{H.}~\bibnamefont{Weigel}},
  \bibinfo{journal}{Int. J. Mod. Phys.} \textbf{\bibinfo{volume}{A11}},
  \bibinfo{pages}{2419} (\bibinfo{year}{1996}).


\bibitem[{\citenamefont{Moussallam}(1993}]{Moussallam:1993jd}
\bibinfo{author}{\bibfnamefont{B.} \bibnamefont{Moussallam}},
  \bibinfo{journal}{Annals Phys.} \textbf{\bibinfo{volume}{225}},
  \bibinfo{pages}{264} (\bibinfo{year}{1993}).



\bibitem[{\citenamefont{Meier}(1996}]{Meier:1996ng}
\bibinfo{author}{\bibfnamefont{F.} \bibnamefont{Meier}}
  \bibnamefont{and}
  \bibinfo{author}{\bibfnamefont{H.}~\bibnamefont{Walliser}},
  \bibinfo{journal}{Phys. Rept.} \textbf{\bibinfo{volume}{289}},
  \bibinfo{pages}{383} (\bibinfo{year}{1997}).


\bibitem[{\citenamefont{Moussallam}(1991}]{Moussallam:1991rj}
\bibinfo{author}{\bibfnamefont{B.} \bibnamefont{Moussallam}}
  \bibnamefont{and}
  \bibinfo{author}{\bibfnamefont{D.}~\bibnamefont{Kalafatis}},
  \bibinfo{journal}{Phys. Lett.} \textbf{\bibinfo{volume}{B272}},
  \bibinfo{pages}{196} (\bibinfo{year}{1991}).




\bibitem[{\citenamefont{Holzwarth}(1995}]{Holzwarth:1995bv}
\bibinfo{author}{\bibfnamefont{G.} \bibnamefont{Holzwarth}}
  \bibnamefont{and}
  \bibinfo{author}{\bibfnamefont{H.}~\bibnamefont{Walliser}},
  \bibinfo{journal}{Nucl. Phys.} \textbf{\bibinfo{volume}{A587}},
  \bibinfo{pages}{721} (\bibinfo{year}{1995}).


\bibitem[{\citenamefont{Weigel:1994gi}(1995}]{Weigel:1994gi}
\bibinfo{author}{\bibfnamefont{H.} \bibnamefont{Weigel}},
\bibinfo{author}{\bibfnamefont{R.} \bibnamefont{Alkofer}},  \bibnamefont{and}
  \bibinfo{author}{\bibfnamefont{H.}~\bibnamefont{Reinhardt}},
  \bibinfo{journal}{Nucl. Phys.} \textbf{\bibinfo{volume}{A582}},
  \bibinfo{pages}{484} (\bibinfo{year}{1995}).





\bibitem[{\citenamefont{Scholtz:1993jg}(1993}]{Scholtz:1993jg}
\bibinfo{author}{\bibfnamefont{F.~G.} \bibnamefont{Weigel}},
\bibinfo{author}{\bibfnamefont{B.} \bibnamefont{Schwesinger}}, \bibnamefont{and}
  \bibinfo{author}{\bibfnamefont{H.~B.}~\bibnamefont{Geyer}},
  \bibinfo{journal}{Nucl. Phys.} \textbf{\bibinfo{volume}{A561}},
  \bibinfo{pages}{542} (\bibinfo{year}{1993}).




\bibitem[{\citenamefont{Halcrow}(1993}]{Halcrow:2015rvz}
\bibinfo{author}{\bibfnamefont{C. J.} \bibnamefont{Halcrow}},
  \bibinfo{journal}{Nucl. Phys.} \textbf{\bibinfo{volume}{B904}},
  \bibinfo{pages}{106} (\bibinfo{year}{2016}).


\bibitem[{\citenamefont{Battye et~al.}(2005)\citenamefont{Battye, Krusch, and Sutcliffe}}]{Battye:2005nx}
\bibinfo{author}{\bibfnamefont{R.~A.} \bibnamefont{Battye}},
  \bibinfo{author}{\bibfnamefont{S.}~\bibnamefont{Krusch}}, \bibnamefont{and}
  \bibinfo{author}{\bibfnamefont{P.~M.} \bibnamefont{Sutcliffe}},
  \bibinfo{journal}{Phys. Lett.} \textbf{\bibinfo{volume}{B626}},
  \bibinfo{pages}{120} (\bibinfo{year}{2005}).



\bibitem[{\citenamefont{Houghton and Magee}(2006)}]{Houghton:2005iu}
\bibinfo{author}{\bibfnamefont{C.}~\bibnamefont{Houghton}} \bibnamefont{and}
  \bibinfo{author}{\bibfnamefont{S.}~\bibnamefont{Magee}},
  \bibinfo{journal}{Phys. Lett.} \textbf{\bibinfo{volume}{B632}},
  \bibinfo{pages}{593} (\bibinfo{year}{2006}).



\bibitem[{\citenamefont{Fortier and Marleau}(2008)}]{Fortier:2008yj}
\bibinfo{author}{\bibfnamefont{J.}~\bibnamefont{Fortier}} \bibnamefont{and}
  \bibinfo{author}{\bibfnamefont{L.}~\bibnamefont{Marleau}},
  \bibinfo{journal}{Phys. Rev.} \textbf{\bibinfo{volume}{D77}},
  \bibinfo{pages}{054017} (\bibinfo{year}{2008}).

\bibitem[{\citenamefont{Foster and Krusch}(2015)}]{Foster:2014vca}
\bibinfo{author}{\bibfnamefont{D.}~\bibnamefont{Foster}} \bibnamefont{and}
  \bibinfo{author}{\bibfnamefont{S.}~\bibnamefont{Krusch}},
  \bibinfo{journal}{Nucl. Phys.} \textbf{\bibinfo{volume}{B897}},
  \bibinfo{pages}{697} (\bibinfo{year}{2015}).



\bibitem[{\citenamefont{Foster and Manton}(2015)}]{Foster:2015cpa}
\bibinfo{author}{\bibfnamefont{D.}~\bibnamefont{Foster}} \bibnamefont{and}
  \bibinfo{author}{\bibfnamefont{N.~S.} \bibnamefont{Manton}},
  \bibinfo{journal}{Nucl. Phys.} \textbf{\bibinfo{volume}{B899}},
  \bibinfo{pages}{513} (\bibinfo{year}{2015}).



\bibitem[{\citenamefont{{Angeli} and {Marinova}}(2013)}]{2013ADNDT..99...69A}
\bibinfo{author}{\bibfnamefont{I.}~\bibnamefont{{Angeli}}} \bibnamefont{and}
  \bibinfo{author}{\bibfnamefont{K.~P.} \bibnamefont{{Marinova}}},
  \bibinfo{journal}{At. Data and Nucl. Data Tables}
  \textbf{\bibinfo{volume}{99}}, \bibinfo{pages}{69} (\bibinfo{year}{2013}).



\bibitem[{\citenamefont{Danilov et~al.}(2009)\citenamefont{Danilov, Belyaeva, Demyanova, Goncharov, and Ogloblin}}]{Danilov:2009zz}
\bibinfo{author}{\bibfnamefont{A.~N.}~\bibnamefont{Danilov}},
  \bibinfo{author}{\bibfnamefont{T.~L.}~\bibnamefont{Belyaeva}},
  \bibinfo{author}{\bibfnamefont{A.~S.}~\bibnamefont{Demyanova}},
  \bibinfo{author}{\bibfnamefont{S.~A.}~\bibnamefont{Goncharov}},
  \bibnamefont{and} \bibinfo{author}{\bibfnamefont{A.~A.}~\bibnamefont{Ogloblin}},
  \bibinfo{journal}{Phys. Rev.} \textbf{\bibinfo{volume}{C80}},
  \bibinfo{pages}{054603} (\bibinfo{year}{2009}).



\bibitem[{\citenamefont{Witten}(1983{\natexlab{a}})}]{Witten1983422}
\bibinfo{author}{\bibfnamefont{E.}~\bibnamefont{Witten}},
  \bibinfo{journal}{Nucl. Phys.} \textbf{\bibinfo{volume}{B223}},
  \bibinfo{pages}{422 } (\bibinfo{year}{1983}{\natexlab{a}}).



\bibitem[{\citenamefont{Witten}(1983{\natexlab{b}})}]{Witten1983433}
\bibinfo{author}{\bibfnamefont{E.}~\bibnamefont{Witten}},
  \bibinfo{journal}{Nucl. Phys.} \textbf{\bibinfo{volume}{B223}},
  \bibinfo{pages}{433 } (\bibinfo{year}{1983}{\natexlab{b}}).



\bibitem[{\citenamefont{Manton and Wood}(2006)}]{Manton:2006tq}
\bibinfo{author}{\bibfnamefont{N.~S.} \bibnamefont{Manton}} \bibnamefont{and}
  \bibinfo{author}{\bibfnamefont{S.~W.} \bibnamefont{Wood}},
  \bibinfo{journal}{Phys. Rev.} \textbf{\bibinfo{volume}{D74}},
  \bibinfo{pages}{125017} (\bibinfo{year}{2006}).



\bibitem[{\citenamefont{Wiringa et~al.}(2000)\citenamefont{Wiringa, Pieper, Carlson, and Pandharipande}}]{Wiringa2000}
\bibinfo{author}{\bibfnamefont{R.~B.} \bibnamefont{Wiringa}},
  \bibinfo{author}{\bibfnamefont{S.~C.} \bibnamefont{Pieper}},
  \bibinfo{author}{\bibfnamefont{J.}~\bibnamefont{Carlson}}, \bibnamefont{and}
  \bibinfo{author}{\bibfnamefont{V.~R.} \bibnamefont{Pandharipande}},
  \bibinfo{journal}{Phys. Rev.} \textbf{\bibinfo{volume}{C62}},
  \bibinfo{pages}{014001} (\bibinfo{year}{2000}).



\bibitem[{\citenamefont{Datar et~al.}(2013)}]{PhysRevLett.111.062502}
\bibinfo{author}{\bibfnamefont{V.~M.} \bibnamefont{Datar}}
  \bibnamefont{et~al.}, \bibinfo{journal}{Phys. Rev. Lett.}
  \textbf{\bibinfo{volume}{111}}, \bibinfo{pages}{062502}
  (\bibinfo{year}{2013}).


  
  
\bibitem[{\citenamefont{Imai et~al.}(2009)\citenamefont{Imai, Aoi, Ong, Sakurai, Demichi et~al.}}]{Imai:2009zza}
\bibinfo{author}{\bibfnamefont{N.}~\bibnamefont{Imai}}
 \bibnamefont{et~al.}, \bibinfo{journal}{Phys. Lett.}
  \textbf{\bibinfo{volume}{B673}}, \bibinfo{pages}{179} (\bibinfo{year}{2009}).
  
  
\bibitem[{\citenamefont{Nagai}(1962)}]{Nagai01041962}
\bibinfo{author}{\bibfnamefont{H.}~\bibnamefont{Nagai}},
  \bibinfo{journal}{Prog. Theor. Phys.} \textbf{\bibinfo{volume}{27}},
  \bibinfo{pages}{619} (\bibinfo{year}{1962}).


\bibitem[{\citenamefont{Epelbaum et~al.}(2014)\citenamefont{Epelbaum, Krebs, L{\"{a}}hde, Lee, Mei{\ss}ner, and Rupak}}]{Epelbaum:2013paa}
\bibinfo{author}{\bibfnamefont{E.}~\bibnamefont{Epelbaum}},
  \bibinfo{author}{\bibfnamefont{H.}~\bibnamefont{Krebs}},
  \bibinfo{author}{\bibfnamefont{T.~A.} \bibnamefont{L{\"{a}}hde}},
  \bibinfo{author}{\bibfnamefont{D.}~\bibnamefont{Lee}},
  \bibinfo{author}{\bibfnamefont{Ulf-G.}~\bibnamefont{Mei{\ss}ner}},
  \bibnamefont{and} \bibinfo{author}{\bibfnamefont{G.}~\bibnamefont{Rupak}},
  \bibinfo{journal}{Phys. Rev. Lett.} \textbf{\bibinfo{volume}{112}},
  \bibinfo{pages}{102501} (\bibinfo{year}{2014}).



\bibitem[{\citenamefont{Esbensen et~al.}(2011)\citenamefont{Esbensen, Tang, and Jiang}}]{Esbensen:2011bg}
\bibinfo{author}{\bibfnamefont{H.}~\bibnamefont{Esbensen}},
  \bibinfo{author}{\bibfnamefont{X.}~\bibnamefont{Tang}}, \bibnamefont{and}
  \bibinfo{author}{\bibfnamefont{C.~L.}~\bibnamefont{Jiang}},
  \bibinfo{journal}{Phys. Rev.} \textbf{\bibinfo{volume}{C84}},
  \bibinfo{pages}{064613} (\bibinfo{year}{2011}).



\bibitem[{\citenamefont{Freer and Fynbo}(2014)}]{Freer:2014qoa}
\bibinfo{author}{\bibfnamefont{M.}~\bibnamefont{Freer}} \bibnamefont{and}
  \bibinfo{author}{\bibfnamefont{H.}~\bibnamefont{Fynbo}},
  \bibinfo{journal}{Prog. Part. Nucl. Phys.} \textbf{\bibinfo{volume}{78}},
  \bibinfo{pages}{1} (\bibinfo{year}{2014}).



\bibitem[{\citenamefont{Wefelmeier}(1937)}]{Wefelmeier1937}
\bibinfo{author}{\bibfnamefont{W.}~\bibnamefont{Wefelmeier}},
  \bibinfo{journal}{Z. Phys.} \textbf{\bibinfo{volume}{107}},
  \bibinfo{pages}{332} (\bibinfo{year}{1937}).



\bibitem[{\citenamefont{Bouten}(1962)}]{Bouten1962}
\bibinfo{author}{\bibfnamefont{M.}~\bibnamefont{Bouten}},
\bibinfo{journal}{Nuovo Cim.} \textbf{\bibinfo{volume}{26}}, 
\bibinfo{pages}{63} (\bibinfo{year}{1962}).



\bibitem[{\citenamefont{Bauhoff et~al.}(1980)\citenamefont{Bauhoff, Schultheis, and Schultheis}}]{Bauhoff:1980zz}
\bibinfo{author}{\bibfnamefont{W.}~\bibnamefont{Bauhoff}},
  \bibinfo{author}{\bibfnamefont{H.}~\bibnamefont{Schultheis}},
  \bibnamefont{and}
  \bibinfo{author}{\bibfnamefont{R.}~\bibnamefont{Schultheis}},
  \bibinfo{journal}{Phys. Rev.} \textbf{\bibinfo{volume}{C22}},
  \bibinfo{pages}{861} (\bibinfo{year}{1980}).



\bibitem[{\citenamefont{Kangasm\"aki et~al.}(1998)}]{PhysRevC.58.699}
\bibinfo{author}{\bibfnamefont{A.}~\bibnamefont{Kangasm\"aki}}
  \bibnamefont{et~al.}, \bibinfo{journal}{Phys. Rev.}
  \textbf{\bibinfo{volume}{C58}}, \bibinfo{pages}{699} (\bibinfo{year}{1998}).



\bibitem[{\citenamefont{Stone}(2005)}]{Stone200575}
\bibinfo{author}{\bibfnamefont{N.}~\bibnamefont{Stone}}, \bibinfo{journal}{At. Data and Nucl. Data Tables} \textbf{\bibinfo{volume}{90}}, \bibinfo{pages}{75} (\bibinfo{year}{2005}).



\bibitem[{\citenamefont{Cheng et~al.}(1974)\citenamefont{Cheng, Goswami, Throop, and McDaniels}}]{PhysRevC.9.1192}
\bibinfo{author}{\bibfnamefont{Y.~T.} \bibnamefont{Cheng}},
  \bibinfo{author}{\bibfnamefont{A.}~\bibnamefont{Goswami}},
  \bibinfo{author}{\bibfnamefont{M.~J.} \bibnamefont{Throop}},
  \bibnamefont{and} \bibinfo{author}{\bibfnamefont{D.~K.}
  \bibnamefont{McDaniels}}, \bibinfo{journal}{Phys. Rev.}
  \textbf{\bibinfo{volume}{C9}}, \bibinfo{pages}{1192} (\bibinfo{year}{1974}).







\bibitem[{\citenamefont{Buck et~al.}(1975)\citenamefont{Buck, Dover, and Vary}}]{Buck:1975zz}
\bibinfo{author}{\bibfnamefont{B.}~\bibnamefont{Buck}},
  \bibinfo{author}{\bibfnamefont{C.}~\bibnamefont{Dover}}, \bibnamefont{and}
  \bibinfo{author}{\bibfnamefont{J.}~\bibnamefont{Vary}},
  \bibinfo{journal}{Phys. Rev.} \textbf{\bibinfo{volume}{C11}},
  \bibinfo{pages}{1803} (\bibinfo{year}{1975}).



\bibitem[{\citenamefont{Ajzenberg-Selove}(1971)}]{AjzenbergSelove19711}
\bibinfo{author}{\bibfnamefont{F.}~\bibnamefont{Ajzenberg-Selove}},
  \bibinfo{journal}{Nucl. Phys.} \textbf{\bibinfo{volume}{A166}},
  \bibinfo{pages}{1} (\bibinfo{year}{1971}).



\bibitem[{\citenamefont{Krutov and Zackrevsky}(1969)}]{0022-3689-2-4-006}
\bibinfo{author}{\bibfnamefont{V.~A.} \bibnamefont{Krutov}} \bibnamefont{and}
  \bibinfo{author}{\bibfnamefont{N.~V.} \bibnamefont{Zackrevsky}},
  \bibinfo{journal}{J. Phys.} \textbf{\bibinfo{volume}{A2}},
  \bibinfo{pages}{448} (\bibinfo{year}{1969}).



\bibitem[{\citenamefont{Allmond et~al.}(2008)\citenamefont{Allmond, Zaballa, Oros-Peusquens, Kulp, and Wood}}]{PhysRevC.78.014302}
\bibinfo{author}{\bibfnamefont{J.~M.} \bibnamefont{Allmond}},
  \bibinfo{author}{\bibfnamefont{R.}~\bibnamefont{Zaballa}},
  \bibinfo{author}{\bibfnamefont{A.~M.} \bibnamefont{Oros-Peusquens}},
  \bibinfo{author}{\bibfnamefont{W.~D.} \bibnamefont{Kulp}}, \bibnamefont{and}
  \bibinfo{author}{\bibfnamefont{J.~L.} \bibnamefont{Wood}},
  \bibinfo{journal}{Phys. Rev.} \textbf{\bibinfo{volume}{C78}},
  \bibinfo{pages}{014302} (\bibinfo{year}{2008}).



\bibitem[{\citenamefont{P{\'{e}}rez and Papenbrock}(2015)}]{Perez:2015tta}
\bibinfo{author}{\bibfnamefont{E.~A.} \bibnamefont{Coello P{\'{e}}rez}}
  \bibnamefont{and}
  \bibinfo{author}{\bibfnamefont{T.}~\bibnamefont{Papenbrock}},
  \bibinfo{journal}{Phys. Rev.} \textbf{\bibinfo{volume}{C92}},
  \bibinfo{pages}{014323} (\bibinfo{year}{2015}).






\end{thebibliography}
\end{document}